\documentclass[11pt,a4paper]{article}
\usepackage{jheppub}
\usepackage{amsmath}

\newcommand{\de}{\mathrm{d}}

\newcommand{\I}{\mathrm{i}}

\newcommand{\cC}{\mathcal{C}}
\newcommand{\cF}{\mathcal{F}}

\newcommand{\cQ}{\mathcal{Q}}

\newcommand{\cI}{\mathcal{I}}
\newcommand{\cM}{\mathcal{M}}

\newcommand{\cN}{\mathcal{N}}
\newcommand{\cE}{\mathcal{E}}

\newcommand{\cR}{\mathcal{R}}

\newcommand{\cJ}{\mathcal{J}}
\newcommand{\cO}{\mathcal{O}}
\newcommand{\cH}{\mathcal{H}}

\newcommand{\cD}{\mathcal{D}}

\newcommand{\nn}{\nonumber}

\newcommand{\IR}{\mathbb{R}}

\newcommand{\IZ}{\mathbb{Z}}
\newcommand{\IN}{\mathbb{N}}

\newcommand{\Tr}{\mbox{Tr}}

\newcommand{\ord}[1]{{\scriptscriptstyle (#1)}}
\newcommand{\Scal}[1]{\Bigl ({#1} \Bigr )}
\newcommand{\scal}[1]{\bigl ({#1} \bigr )}

\def\inv{{ \scriptscriptstyle \rm{\mbox{\tiny-1}}}}

\def\bea{\begin{eqnarray}}
\def\eea{\end{eqnarray}}
\def\be{\begin{equation}}
\def\ee{\end{equation}}
\def\ba{\begin{align}}
\def\ea{\end{align}}
\def\bse{\begin{subequations}}
\def\ese{\end{subequations}}

\def\vI{v^{\mbox{-}1{\rm t}}}

\def\RN{{\rm R.N.}}

\def\vq{{ q}}
\def\vp{{ p}}
\def\tr{{\rm tr\, }}

\title{Exact $\nabla^4\cR^4$ couplings and helicity supertraces}

\preprint{CERN-TH-2016-224, CPHT-RR046.102016}

\author[a]{Guillaume Bossard,}
\author[b,c]{Boris Pioline,} 

\affiliation[a]{Centre 
de Physique Th\'eorique, Ecole Polytechnique, Universit\'e Paris-Saclay, 91128 Palaiseau Cedex, France}

\affiliation[b]{Laboratoire de Physique Th\'eorique et Hautes
Energies, CNRS UMR 7589, \\
Universit\'e Pierre et Marie Curie,
4 place Jussieu, 75252 Paris cedex 05, France} 

\affiliation[c]{CERN, Theoretical Physics Department, 1211 Geneva 23, Switzerland}
 
\emailAdd{guillaume.bossard@polytechnique.edu}
\emailAdd{boris.pioline@cern.ch}

\abstract{In type II string theory compactified on a $d$-dimensional torus $T^d$
down to $D=10-d$ dimensions, the $\cR^4$ and $\nabla^4\cR^4$ four-graviton couplings 
are known exactly, for all values of the moduli, in terms of certain Eisenstein series of the U-duality group $E_{d}(\IZ)$. 
In the limit where one circle in the torus becomes large, these couplings are expected to reduce
to their counterpart in dimension $D+1$, plus threshold effects and exponentially suppressed
corrections corresponding to BPS black holes in dimension $D+1$ whose worldline
winds around the circle. By combining the weak coupling and large radius limits, we 
determine these exponentially suppressed corrections exactly, 
and demonstrate that the contributions of 1/4-BPS black holes 
to the $\nabla^4\cR^4$ coupling are proportional to the appropriate helicity supertrace.
Mathematically, our results provide the complete Fourier expansion of the next-to-minimal theta
series of $E_{d+1}(\IZ)$ with respect to the maximal parabolic subgroup with Levi component 
$E_{d}$ for $d\leq 6$, and the complete Abelian part of the Fourier expansion
of the same for $d=7$.}

\begin{document}

\maketitle

\section{Introduction}

The lowest four-graviton couplings  in the effective action of string vacua with maximal 
supersymmetry have been under intense scrutiny since \cite{Green:1997tv}, as they provide 
one of the few examples of non-trivial observables in string theory which are computable exactly, beyond perturbation theory.  Indeed, by combining invariance under U-duality, supersymmetric Ward identities,  and perturbative computations at low order, it was shown  that
the $\cR^4$ and $\nabla^4\cR^4$ couplings in type II string theory compactified on a torus $T^d$ 
down to $D=10-d$ dimensions are given by specific automorphic functions of the moduli fields known as maximal parabolic Eisenstein series \cite{Green:1997di,Kiritsis:1997em,Green:1997as,Pioline:1998mn,
Green:1998by,Obers:1999um,Green:1999pu,Basu:2007ru,Green:2010wi,Pioline:2010kb,Green:2010kv,Green:2011vz,Bossard:2014lra,Bossard:2014aea}. More precisely, the $\cR^4$ and $\nabla^4\cR^4$ are proportional to the minimal and next-to-minimal theta series of the U-duality 
group $G_{D}(\IZ)\equiv E_{d+1}(\IZ)$, which arise
from the maximal parabolic Eisenstein series $\mathcal{E}^{G_D}_{s\Lambda_1}$ at special values $s=3/2, 5/2$ of the parameter $s$.
Besides reproducing the known perturbative contributions at weak string coupling, and predicting the vanishing of all higher order perturbative contributions, these automorphic functions also display an infinite series of instanton corrections from BPS branes wrapping supersymmetric cycles in $T^d$, providing invaluable probes into the non-perturbative regime of string theory. Mathematically, these 
instanton corrections correspond to the Fourier coefficients of $\mathcal{E}^{G_D}_{s\Lambda_1}$
with respect to the maximal parabolic subgroup $P_1$ with Levi subgroup $\IR^+\times Spin(d,d)$.
Despite being also protected by supersymmetry, the $\nabla^6\cR^4$ coupling is not a parabolic Eisenstein series (or residue thereof) but a more complicated type of automorphic function \cite{Green:2005ba,Basu:2007ck,Green:2014yxa,Pioline:2015yea,Bossard:2015foa}, and its instantonic corrections have yet to be fully analyzed .  

At the same time, these protected couplings in dimension $D$ can serve as useful
book-keeping device for precision counting of black hole micro-states in dimension 
$D+1$ \cite{Gunaydin:2005mx,Pioline:2006ni}.
Indeed, in the limit where the radius $R$ of one circle in $T^d$ becomes very large,  effective couplings in dimension $D$ are expected to reduce to the same couplings in dimension $D+1$, up to power-like threshold effects \cite{Green:1999pu,Green:2006gt} and, more importantly for us, exponentially suppressed corrections from Euclideanized $D+1$-dimensional black holes winding around the circle (in addition, when $D=3$ there are also further exponentially suppressed  contributions from Taub-NUT instantons). In particular, the $\cR^4$,  $\nabla^4 \cR^4$  and
 $\nabla^6 \cR^4$ couplings are expected to receive contributions from 1/2-BPS, 1/4 and 1/8-BPS black holes in dimension $D+1$, respectively, weighted by the corresponding BPS indices. 
Indeed,  general mathematical results on the wave-front set of automorphic 
forms \cite{matumoto1987whittaker,miller2012fourier} 
can be used to show that the Fourier coefficients of the minimal and next-to-minimal theta series with respect to the maximal parabolic subgroup $P_{d+1}$ with Levi $\IR^+ \times E_{d(d)}$ vanish unless the charge vector satisfies the corresponding 1/2-BPS or 1/4-BPS constraint \cite{Green:2011vz,Pioline:2010kb}. These constraints also follow from the tensorial differential equations satisfied by these couplings, which in turn are consequences of supersymmetry Ward identities \cite{Bossard:2014lra,Bossard:2014aea,Bossard:2015uga}. In the case of the $\nabla^6\cR^4$ coupling,
the analysis of \cite{Bossard:2015uga,Bossard:2015foa} confirms that the wave-front set
of $\cE_{\scalebox{0.6}{$(0,1)$}}^{\scalebox{0.6}{$(D)$}}$ for $D=3$ coincides with the nilpotent orbit supporting 1/8-BPS 
black holes \cite{Bossard:2009at}, 
whereas it suggests that this connection may be lost in $D=4$. The relation between the summation measure for instantons in dimension $D=3$ and BPS index in dimension $D+1=4$ is also well established in the case of supersymmetric gauge theories \cite{Dorey:2000qc,Gaiotto:2008cd,Alexandrov:2014wca}, and was demonstrated recently in the analogous case of $F^4$ and $D^2 F^4$ couplings in string vacua with half-maximal supersymmetry \cite{Bossard:2016zdx}. Our main goal in this paper will be to show that this expectation is also borne out for the 
$\nabla^4 \cR^4$ couplings in type II string theory compactified on $T^{10-D}$ with $D=3,4,5$: namely, that the
summation measure for 1/4-BPS instantons of charge $Q$ in dimension $D$ reproduces the 
expected helicity supertrace $\Omega_{12}(Q)$ which counts (with signs) 1/4-BPS black hole states in $D+1$ dimensions.
In the process, we shall also obtain the complete Fourier expansion of the next-to-minimal theta
series for $E_{d+1(d+1)}$ with respect to the parabolic subgroup $P_{d+1}$ for $d\leq 6$, or its Abelian part in the case of $d=7$. 

The outline of this note is as follows. In Section \ref{sec_bps}, we review well known facts about
BPS states and helicity supertraces in toroidal compactifications of type II string theories, In Section
\ref{sec_decomp}, we study the decompactification limit of the $\nabla^4\cR^4$ coupling in dimension
$D$, by first considering the large radius limit of the known perturbative contributions, and then
covariantizing the result under U-duality. In Section \ref{sec_latsum}, we provide an alternative
derivation of this result by representing the maximal parabolic Eisenstein series as an Epstein-type sum over an $E_d(\IZ)$ invariant lattice subject to certain quadratic constraints. In particular
for $d=8$ we construct a lattice in the Lie algebra of $E_{8(8)}$ which is invariant under the
Chevalley group $E_8(\IZ)$ and has a single orbit of primitive rank-one elements. 
We conclude with a brief discussion in Section \ref{sec_discuss}. Appendix A contains computations of decompactification
limits of certain modular integrals which appear in the perturbative expansion of the $\nabla^4\cR^4$
coupling, appendix B studies the tensorial differential equations satisfied by the Fourier coefficients, and appendix C gives some details on the construction of certain invariant lattices 
under the Chevalley groups $E_{d+1(d+1)}(\IZ)$.

\section{BPS degeneracies and helicity supertraces\label{sec_bps}}

Before studying the Fourier expansion of the minimal theta series, it is useful to review some known results about type II string theory compactified on $T^6$. Recall it is described at low energy by $\cN=8$ supergravity, including 70 scalar fields valued in the coset space $E_{7(7)}/SU(8)$, 28 Maxwell fields and their magnetic duals. Electromagnetic charges take values in a 56-dimensional lattice, equipped with an
integer symplectic pairing $\langle \Gamma,\Gamma'\rangle$ invariant under the action of 
$E_{7}(\IZ)$. Since the charge lattice transforms
as a module with highest weight $\Lambda_7$ under $E_{7}(\IZ)$, where $\Lambda_7$ is the
fundamental weight associated to the 7-th node in the Dynkin diagram of $E_7$ using Bourbaki's
labeling, we shall denote this lattice by $M^{E_7}_{\Lambda_7} \cong  \IZ^{56}$ 
(see Appendix \ref{sec_e8lat} for some details on the construction of $M^{E_7}_{\Lambda_7}$ and other lattices).

The supersymmetry algebra  $\{Q^i_\alpha,Q^j_\beta\}= \varepsilon_{\alpha\beta} Z^{ij}$ 
includes a complex antisymmetric $8\times 8$ matrix $Z^{ij}=\overline{Z_{ij}}$ known as the central charge matrix, 
which depends on the moduli in $E_{7(7)}/SU(8)$. Generic supersymmetric multiplets constructed
out of a spin $j$ representation have $2^{15}D_j$ bosonic states and as many fermionic states, with $D_j=2j+1$. Half-BPS states occur when all skew-eigenvalues of $Z^{ij}$ are equal in absolute value and the sum of their phases vanish, and have $2^7D_j$ bosonic states and mass
\be
\label{mass12}
\cM_{1/2}(\Gamma)= |Z(\Gamma)|^2 = Z_{ij}(\Gamma) Z^{ij}(\Gamma)\ .
\ee
1/4-BPS states occur when the absolute values of the skew-eigenvalues of $Z^{ij}$ are equal in pairs and the sum of their phases vanish, and have $2^{11}D_j$ bosonic states and mass
\be
\label{mass14}
\cM_{1/4}(\Gamma) = \sqrt{ |Z(\Gamma)|^2+ 2 \sqrt{\Delta(\Gamma)}}\ ,
\ee
where
\be
\Delta(\Gamma) = 2 Z_{ij}(\Gamma)  Z^{jk}(\Gamma) Z_{kl}(\Gamma) Z^{li}(\Gamma) -\frac{1}{4} (Z_{ij}(\Gamma) Z^{ij}(\Gamma) )^2\ .
\ee
1/8-BPS states occur when at least one skew-eigenvalue of $Z^{ij}$ does not vanish and have $2^{13}D_j$ bosonic states. Their mass formula will not be needed in this paper. 

It was shown in \cite{Ferrara:1997ci} that the 1/2-BPS  condition on the central charge matrix $Z_{ij}(\Gamma)$ for black hole solutions is equivalent
to the condition
\be
\label{bps12cond}
1/2\mbox{-BPS}: \ \Gamma\times\Gamma=0\ ,\qquad 
\ee
on the charge vector $\Gamma\in\IZ^{56}$, where $\Gamma\times\Gamma$ denotes the projection
of the symmetric tensor product ${\bf 56}\otimes_s {\bf 56}$ on the adjoint representation 
${\bf 133}$ of  $E_{7(7)}$.
The condition \eqref{bps12cond} defines a 28-dimensional cone $\cC_{1/2}$ inside $\IR^{56}$, isomorphic to the homogeneous space $E_{7(7)} / (E_{6(6)}\ltimes \IR^{27})$, where the 
denominator is the stabilizer of a non-zero 1/2-BPS charge vector \cite{Ferrara:1997uz}. 
We shall refer to vectors $\Gamma$ such that
$\Gamma\times\Gamma=0$ as `rank-one' vectors.
Similarly, the 1/4-BPS condition 
on $Z_{ij}(\Gamma)$ is equivalent
to the condition
\be
\label{bps14cond}
1/4\mbox{-BPS}:\ I'_4(\Gamma)=0\ ,\quad \Gamma\times\Gamma\neq 0
\ee
on the charge vector, where $I'_4(\Gamma)$ is the projection of  ${\bf 56}\otimes_s {\bf 56}$ on 
the ${\bf 56}$
representation of  $E_{7(7)}$. Equivalently, $I'_4(\Gamma)$ is the gradient of 
the quartic polynomial
\bea
I_4(\Gamma) &=& 16 Z_{ij}(\Gamma)  Z^{jk}(\Gamma) Z_{kl}(\Gamma) Z^{li}(\Gamma) -4 (Z_{ij}(\Gamma) Z^{ij}(\Gamma) )^2
\\
&& +\frac{1}{6} \varepsilon_{ijklmnpq} Z^{ij}(\Gamma) Z^{kl}(\Gamma) Z^{mn}(\Gamma) Z^{pq}(\Gamma)
+\frac{1}{6}Ê\varepsilon^{ijklmnpq} Z_{ij}(\Gamma) Z_{kl}(\Gamma) Z_{mn}(\Gamma) Z_{pq}(\Gamma) \ , \nn 
\eea
which is invariant under the action of $E_{7(7)}$ on $\IR^{56}$ and independent of the moduli. 
The conditions \eqref{bps14cond} define a 45-dimensional cone $\cC_{1/4}$ in $\IR^{56}$, isomorphic to
the homogenous space $E_{7(7)}/[Spin(6,5)\ltimes \IR^{32+1}]$, where the denominator is
the stabilizer of a non-zero 1/4-BPS charge vector \cite{Ferrara:1997uz}. 
We shall refer to vectors $\Gamma$ satisfying
\eqref{bps14cond} as `rank-two' vectors. It is worth noting that the condition $\Gamma\times\Gamma=0$ implies $I'_4(\Gamma)=0$, which itself implies $I_4(\Gamma)=0$. In contrast, generic 1/8-BPS states have $I_4(\Gamma)\neq 0$.

In general, to count BPS states preserving a certain fraction of supersymmetries, it is useful to consider the helicity supertrace \cite{Ferrara:1981qd,Kiritsis:1997hj}\footnote{In this section we closely follow Appendices E and G in \cite{Kiritsis:1997hj}, except for a change of
normalization.}
\be
\Omega_{n}(\Gamma) = \frac{(-1)^{n/2}}{n!}\, \Tr'_\Gamma (-1)^{2J_3} (2J_3)^n\ ,
\ee 
where $\Tr'_\Gamma$ denotes the trace in the superselection sector with electromagnetic charge $\Gamma\in\IZ^{56}$, after factoring out the bosonic center of mass degrees of freedom, and the helicity $J_3$ is the generator of rotations around a fixed axis. The traces $\Omega_n$ are most easily obtained from the `helicity generating function, or `refined index' $\Omega(\Gamma,y)$,
\be
\Omega_{n}(\Gamma) = \frac{(-1)^{n/2}}{n!}\, \left(y\partial_y\right)^n \Omega(\Gamma,y)\vert_{y=1}\ ,\quad 
\Omega(\Gamma,y) = \Tr_\Gamma (-1)^{2J_3} y^{2J_3}\ .
\ee
By parity, $\Omega_n$ vanishes unless $n$ is even. For $\cN=8$ supersymmetry in $D=4$, $\Omega_{n\leq 6}$ vanishes on all supersymmetry multiplets, while $\Omega_8$, $\Omega_{10}$, 
$\Omega_{12}$, $\Omega_{14}$ vanish on all multiplets except on
those preserving $1/2, 1/4$ or $1/8$ of the supersymmetry of the vacuum, respectively.

In this paper we shall restrict attention to states which preserve at least $1/4$ of
the $\cN=8$ supersymmetries, so it suffices to consider the traces $\Omega_8$, $\Omega_{10}$, $\Omega_{12}$. 
A  1/2-BPS spin $j$ multiplet contributes
\be
\Omega_8=(-1)^{2j} D_j\ ,\quad \Omega_{10}=\frac16(-1)^{2j+1} D_j(D_j^2+1)\ ,\quad
\Omega_{12}=\frac{1}{360} (-1)^{2j} D_j(3D_j^4+10 D_j^2+6)\ ,
\ee
while a 1/4-BPS spin $j$  contributes
\be
\Omega_8= \Omega_{10}=0 ,\quad \Omega_{12}=(-1)^{2j+1} D_j\ .
\ee

In the perturbative spectrum of type II string compactified on $T^6$, all states are neutral
under the Ramond-Ramond (RR) gauge fields and carry no magnetic charge under the 
Neveu--Schwarz gauge fields, so the electromagnetic charge vector $\Gamma=(Q,0,0)$ 
has support only on the first term in the decomposition of the charge lattice $\IZ^{56}=\IZ^{6,6}\oplus \IZ^{32}\oplus \IZ^{6,6}$ under the T-duality group 
$Spin(6,6)\subset E_{7(7)}$. The vector $Q=(m_i, w^i)\in\IZ^{6,6}$ encodes the momenta $m_i$ and winding
numbers $w^i$ of the string along $T^6$. For such lattice vectors, the quantities $|Z(\Gamma)|^2$
and $\Delta(\Gamma)$ entering the mass formulae \eqref{mass12} and \eqref{mass14} become
\be
\label{ZtopLR}
|Z(\Gamma)|^2 = g_D^{\frac{4}{d-8}}[p_L^2(Q) +p_R^2(Q)]\ ,\quad \Delta(\Gamma)=
\frac14 g_D^{\frac{8}{8-d}}\, \left(
p_L^2(Q) -p_R^2(Q) \right)^2
\ee
where $p_L(Q)$ and $p_R(Q)$ are the projections of the electric charge vector $Q\in \IZ^{6,6}$ on the
spacelike $6$-plane and its orthogonal complement inside $\IR^{12}$ labelled by the Narain moduli
$O(6,6)/[O(6)\times O(6)]$, such that the norm on the even self-dual Narain lattice is 
$(Q,Q)= p_L^2(Q) -p_R^2(Q) = 2 m_i w^i$; and $g_D$ is the string coupling constant in $D$ dimensions.

A straightforward light-cone quantization of the type II string leads to \cite{Kiritsis:1997hj}
\be
\Omega(\rho_2,y) \equiv
\sum_{Q\in \IZ^{6,6}} \Omega(Q,y)\, e^{-\pi \rho_2 ( p_L^2(Q) +p_R^2(Q))} =
\int_0^1 \de\rho_1\, \Gamma_{6,6}(\rho) \left| \frac{2\eta^3(\rho) \, \sin\pi v}{\theta_1(\rho,v)} \frac{ \theta_1^4(\rho,v/2)}{\eta^{12}(\rho)}\right|^2\ ,
\ee
where $q=e^{2\pi\I\rho}, y=e^{2\pi\I v}$, $\eta(\rho)$ and $\theta_1(\rho,v)$ are the Dedekind and Jacobi theta functions, and $\Gamma_{6,6}=\rho_2^3 \sum_{Q\in \IZ^{6,6}} q^{\frac12 p_L^2(Q)} \bar q^{\frac12 p_R^2(Q)}$ is the Siegel--Narain theta series. 
Taking derivatives with respect to $v$ and setting $v=0$, one readily finds
\bea
\label{Omrho}
\Omega_8(\rho_2)&=&\int d\rho_1\Gamma_{6,6}\ ,\qquad \Omega_{10}(\rho_2)=-\frac13\int d\rho_1\Gamma_{6,6}\ ,\nn \\
\Omega_{12}(\rho_2)&=&\int d\rho_1\left(\frac{19}{360} + \frac{E_4+\bar E_4-2}{240}\right)\Gamma_{6,6} \ , 
\eea
where  $E_4=1+240 \sum_{N=1}^{\infty} \sigma_3(N) q^N$ is the holomorphic Eisenstein series of weight $4$, and $\sigma_s(N)=\sum_{d|N} d^s$ is the
standard divisor function.

The  result for $\Omega_8(\rho_2)$ shows that unpaired 1/2-BPS states exist only in the left and right-moving ground state of the string, such that $m_i w^i=0$, and that $\sum_j (-1)^{2j} D_j=1$ for these states. In fact, one can show that there are no cancellations in this sum, and that there exists a single 1/2-BPS multiplet with spin $j=0$ and mass $\cM(Q)=\sqrt2 |p_L(Q)|=\sqrt2 |p_R(Q)|$ for any vector $Q\in\IZ^{6,6}_*\equiv \IZ^{6,6}\backslash\{0\}$ such that  $(Q,Q)=0$. Indeed the contribution of this multiplet is apparent in $\Omega_{10}$ and in the first term of $\Omega_{12}$. The result for $\Omega_{12}(\rho_2)$ on the other hand shows that there are unpaired 1/4-BPS states when the string is either in the left or in the right-moving ground state, with mass $\cM(Q)=\sqrt2 |p_L(Q)|$ or $\sqrt2 |p_R(Q)|$, respectively,
and that $\sum_j (-1)^{2j} D_j= \sigma_3[\tfrac12(Q,Q)]$. In contrast with the 1/2-BPS case, there are in fact huge cancellations in this sum, since the number of
bosonic states with excitation level $N$ on left-moving side and in the ground state on the right-moving  side (or vice-versa) is given by the $N$-th Fourier coefficient of $\theta_2^4/\eta^{12}$, which grows 
exponentially with $N$ \cite{Dabholkar:1989jt}. In contrast, the divisor function $\sigma_3(N)$ grows like $N^3$, which indicates that most perturbative 1/4-BPS states can pair up
and decay when the string coupling becomes non-zero. 
The computation of $\Omega_{14}(\rho_2)$ shows that there are no unpaired 1/8-BPS states in the perturbative spectrum,
in fact all states with non-zero excitation level on both the left and right-hand sides 
lie in long multiplets which only
contribute to $\Omega_n$ with $n\geq 16$ \cite{Kiritsis:1997hj}. 
 
At the non-perturbative level, it is now well-known that there exists many 1/8-BPS states with arbitrary electromagnetic charge $\Gamma\in \IZ^{56}$, corresponding to bound states of perturbative strings, D-branes, NS5-branes and Kaluza--Klein monopoles wrapped on $T^6$. 
The helicity supertrace $\Omega_{14}(\Gamma)$ associated to these 1/8-BPS states was determined
for arbitrary $\Gamma$ in \cite{Shih:2005qf,Pioline:2005vi,Sen:2008sp,Sen:2009gy}, and grows exponentially as $e^{\pi\sqrt{I_4(\Gamma)}}$ as $I_4(\Gamma)$ becomes large and positive, in agreement with the Bekenstein--Hawking entropy of 1/8-BPS black holes 
in $\cN=8$ supergravity \cite{Kallosh:1996uy}. This  helicity supertrace would be relevant for analyzing the decompactification limit of $D^6 \cR^4$ couplings, but it will not be needed in this work. 
In contrast, 1/4-BPS and 1/2-BPS states necessarily have $I_4(\Gamma)=0$, and do not correspond to any regular, classical black hole (the Riemann tensor in the spherically symmetric BPS solution with charge $\Gamma$ is not bounded outside the point-like horizon, although this problem may get resolved for 1/4-BPS black holes when taking into account higher-derivative corrections \cite{Dabholkar:2004yr,Sinha:2006yy}). It is natural to expect (and will be vindicated by our analysis) that the only 1/2-BPS and 1/4-BPS states are those which lie in the same duality orbit as the perturbative BPS states in 
type II string theory. Thus, we assume that for a primitive charge $\Gamma$ (with unit greatest common divisor, gcd$(\Gamma)=1$),
\be
 \label{om8}
\Omega_8(\Gamma)=\begin{cases}
1 & (\Gamma\times \Gamma=0) \\
0 & (\Gamma\times \Gamma\neq 0)
\end{cases}
\ee
 \be
 \label{om12}
\Omega_{12}(\Gamma) =
\begin{cases}
\frac{19}{360} & (\Gamma\times \Gamma=0) \\
 \sigma_3[{\rm gcd}(\Gamma\times\Gamma)] & (I'_4(\Gamma)=0, \Gamma\times \Gamma\neq 0)\\
 0 & (I'_4(\Gamma)\neq 0)
 \end{cases}
\ee
When $\Gamma$ is non-primitive, {\it i.e.} $\Gamma=n\Gamma_0$ with $\Gamma_0\in\IZ^{56}$, it is expected that there exist a single threshold bound state for any $n$ when $\Gamma\times\Gamma=0$
(in line with the fact that D0-branes are Kaluza--Klein modes of the eleven-dimensional graviton), while
the existence of threshold bound states in the 1/4-BPS or 1/8-BPS cases is as far as we know an open problem.

In this section, we have focussed on BPS states and helicity supertraces in type II string compactified on $T^6$, however a similar classification exists in type II string
on $\IR^{1,D-1} \times T^d$ down to $D>4$:
the charge lattice $M^{E_{d+1}}_{\Lambda_{d+1}}$ is a $\IZ$-module of $E_{d+1}(\IZ)$ 
with highest weight $\Lambda_{d+1}$ in Bourbaki's labelling of the fundamental weights; 1/2-BPS states exist for any $D\geq 4$, and are such that the projection of $\Gamma \otimes_s \Gamma$ on the module with highest weight $\Lambda_1$ vanishes (a condition which we continue
to write as $\Gamma\times \Gamma=0$); 1/4-BPS states exist for $8\geq D\geq 4$,  have generic charge vectors for $8\geq D\geq 6$ and satisfy a cubic
condition $I_3(\Gamma)=0$ for $D=5$, which by abuse of notation we continue
to write as $I'_4(\Gamma)=0$.  Analogues of the helicity supertraces $\Omega_{8}$ and
$\Omega_{12}$ can be defined, for which the results \eqref{om8} and \eqref{om12} continue to hold.

\section{Decompactification limit of $\nabla^4\cR^4$ couplings\label{sec_decomp}}

As mentioned in the introduction, the coefficients of the $\cR^4$ and $\nabla^4\cR^4$  couplings in the
low energy effective action of type II strings on $\IR^{1,D-1}\times T^d$ with $D=10-d$, which we
denote by $\cE_{\scalebox{0.6}{$(0,0)$}}^{\scalebox{0.6}{$(D)$}}$ and $\cE_{\scalebox{0.6}{$(1,0)$}}^{\scalebox{0.6}{$(D)$}}$ following  \cite{Green:2010wi}, are given by maximal parabolic Eisenstein series of the U-duality group $E_{d+1}(\IZ)$ at special values of the parameter $s$. More precisely,\footnote{For $D=8$, the U-duality
group is no longer simple and $\cE_{(0,0)}^{(8)}$ involve a sum of regularized Eisenstein series for
$SL(2,\IZ)$ and $SL(3,\IZ)$ \cite{Kiritsis:1997em}, while $\cE_{(1,0)}^{(8)}$ involves products of Eisenstein series for $SL(2,\IZ)$ and $SL(3,\IZ)$ \cite{Basu:2007ru}. For $D=6,7$, $\cE_{\scalebox{0.6}{$(1,0)$}}^{\scalebox{0.6}{$(D)$}}$ involves sums of the regularized Eisenstein series $\widehat\cE^{E_{d+1}}_{\frac{5}{2}\Lambda_1}$ and $\widehat\cE^{E_{d+1}}_{\frac{d+2}{2}\Lambda_{d+1}}$ \cite{Green:2010wi}.}
\be
\label{eisr4}
\cE_{(0,0)}^{(D)} = 2\zeta(3)\, \cE^{E_{d+1}}_{\frac32 \Lambda_1}(g) =
4\pi\, \xi(d-2)\,  \cE^{E_{d+1}}_{\frac{d-2}{2} \Lambda_{d+1}}(g)
\qquad (D<8)
\ee
\be
\label{eisd4r4}
\cE_{(1,0)}^{(D)} = \zeta(5)\, \cE^{E_{d+1}}_{\frac52 \Lambda_1}(g)
=8\pi\, \xi(4)\,\xi(d+2)\, \cE^{E_{d+1}}_{\frac{d+2}{2} \Lambda_{d+1}}(g) \quad (D<6)
\ee
where we denote by $\cE^{G}_{\lambda}(g)$ the maximal parabolic Eisenstein series with infinitesimal
character $2\lambda-\rho_G$ evaluated at $g\in G/K$, $\rho_G$ being the Weyl vector, and $\xi(s)=\pi^{-s/2}\Gamma(s/2) \zeta(s)$ is the completed Riemann zeta function, invariant under $s\mapsto 1-s$. In writing the second equality in \eqref{eisr4} and \eqref{eisd4r4}, we used the functional equation for Langlands--Eisenstein series (see e.g. \cite{Fleig:2015vky} for an extensive introduction to 
Eisenstein series, or \cite[\S A]{Pioline:2015yea} for a telegraphic account). As mentioned in the
introduction, the Eisenstein series $\cE^{E_{d+1}}_{s\Lambda_1}$ and $ \cE^{E_{d+1}}_{s\Lambda_{d+1}}$ for the special values of the parameter $s$ appearing in \eqref{eisr4} and \eqref{eisd4r4}
are singled out physically by supersymmetric Ward identities, and mathematically by the fact that they are attached to the minimal and next-to-minimal nilpotent orbits of $E_{d+1(d+1)}$, respectively. Accordingly, we refer to them as the minimal and next-to-minimal theta series, respectively. Let us introduce a few important mathematical concepts to explain what this means.

In general, automorphic forms under $G(\IZ)$ can be viewed as vectors in 
certain automorphic representations of $G(\IR)$. The automorphic representation characterizes
the set of differential equations that the automorphic form satisfies, in terms of a certain ideal in the enveloping algebra of $\mathfrak{g}$. The variety associated to this ideal can be
shown to be the closure of a unique special nilpotent adjoint orbit $\cO$ in $\mathfrak{g}$. The Gelfand--Kirillov dimension of an automorphic representation is half the dimension of the nilpotent 
orbit $\cO$. For any parabolic subgroup $P \cong LU$ of $G$, with unipotent radical $U$ and
Levi subgroup $L$,  the dimension of the maximal Abelian subgroup of $U$ for which the Fourier coefficients are non-zero is bounded from above by the Gelfand--Kirillov dimension. The  
values of the parameter $s$ appearing in \eqref{eisr4} and \eqref{eisd4r4} are special in that the
orbits associated to the corresponding automorphic representations are the smallest nilpotent orbit
$\cO_{\rm min}$
in the case of  $\cE_{\scalebox{0.6}{$(0,0)$}}^{\scalebox{0.6}{$(D)$}}$ \cite{Pioline:2010kb}, and the next-to-smallest orbit 
$\cO_{\rm ntm}$  in the case 
of $\cE_{\scalebox{0.6}{$(1,0)$}}^{\scalebox{0.6}{$(D)$}}$ \cite{Green:2010wi,Green:2011vz}. The Gelfand--Kirillov dimensions of the minimal and next-to-minimal orbits for the groups $E_{d+1(d+1)}$ are tabulated in Table 1. Although this was already conjectured before by Ginzburg, a recent theorem \cite{gomez2016whittaker}  shows that there is no cuspidal representation attached to the minimal and the next-to-minimal nilpotent orbits for $E_7$ and $E_6$, strengthening the result of \cite{miller2012fourier} and proving that the functions \eqref{eisr4} and  \eqref{eisd4r4} are indeed uniquely determined by U-duality and supersymmetry Ward identities.

Nilpotent orbits also appear in the classification of spherically symmetric black holes \cite{Gunaydin:2005mx,Bossard:2009at}. For example, 1/2-BPS black holes in $D=4$ dimensions with charge $\Gamma\in \cC_{1/2}$ (and vanishing NUT charge) are parametrised by a 
Lagrangian submanifold of the minimal $E_{8(8)}$ nilpotent orbit defined by its intersection with the complement of $\mathfrak{so}^*(16)$ in $\mathfrak{e}_{8(8)}$. Similarly, 1/4-BPS black hole solutions  with charge in $\Gamma\in \cC_{1/4}$ are parametrised by a Lagrangian of the next-to-minimal $E_{8(8)}$ nilpotent orbit \cite{Bossard:2009at}. This fact generalises to higher dimensions, and implies that the Gelfand--Kirillov dimension of the minimal representation
of $E_{d+2(d+2)}$ coincides with $\dim \cC_{1/2}$ for $D>4$ and with $\dim \cC_{1/2}+1$ for $D=4$ (where the extra $1$ originates from the NUT charge); similarly, the Gelfand--Kirillov dimension of the next-to-minimal representation of $E_{d+2(d+2)}$ coincides with $\dim \cC_{1/4}$ for $D>4$ and with $\dim \cC_{1/4}+1$ for $D=4$ (see Table 1). Moreover, the stabilizer of a 1/2-BPS charge vector in dimension $D$ is by construction the parabolic subgroup $P_{d+1}$ of $E_{d+1(d+1)}$ associated to the weight $\Lambda_{d+1}$ whose Levi subgroup (modulo the center) happens to coincide with the U-duality group $E_{d(d)}$ in dimension $D+1$. These observations will play an important role in the following.

\begin{table}
$$\begin{array}{|c|c||c|c||c|c|c|}
\hline
D & E_{d+1(d+1)} & \tfrac12\dim \cO_{\rm min} & \tfrac12\dim \cO_{\rm ntm} 
& \dim M^{E_{d+1}}_{\Lambda_{d+1}} 
& \dim \cC_{1/2} &  \dim \cC_{1/4}  \\
\hline
3  & E_{8(8)} & 29 & 46& 248 & 58 & 92 \\
4  & E_{7(7)} & 17 & 26 & 56 & 28 & 45\\
5 & E_{6(6)} &  11 & 16 &  27 & 17 & 26\\
6 & Spin(5,5) & 7 & 10 & 16 & 11 & 16 \\
\hline
\end{array}
$$
\caption{U-duality group in type II string theory on $\IR^{1,D-1}\times T^{10-D}$, Gelfand--Kirillov dimension of the minimal and next-to-minimal unipotent representations, dimension of the charge lattice and of the homogeneous cones $\cC_{1/2}$ and $\cC_{1/4}$ 
of 1/2-BPS and 1/4-BPS charges.}
\end{table}

Our goal in the remainder of this section is to determine the behavior of the $\nabla^4\cR^4$ coupling
$\cE_{\scalebox{0.6}{$(1,0)$}}^{\scalebox{0.6}{$(D)$}}$ in the limit where one circle inside $T^d$ becomes very large, keeping track
of exponentially suppressed corrections. In mathematical terms, we shall compute the Fourier
coefficients of $\cE_{\scalebox{0.6}{$(1,0)$}}^{\scalebox{0.6}{$(D)$}}$ with respect to the parabolic subgroup $P_{d+1}$. Our main interest will be in the $D=3$ case, but our method is general and we shall keep $d$ arbitrary for the most part. In order to compute the Fourier expansion of $\cE_{\scalebox{0.6}{$(1,0)$}}^{\scalebox{0.6}{$(D)$}}$, our strategy will be to simultaneously take the decompactification and weak coupling limits, 
as in \cite[\S 2.2.1]{Pioline:2015yea}, and recover the generic Fourier coefficients with respect to $P_{d+1}$ by covariantizing the large radius limit of the Fourier expansion of the perturbative contributions.
Recall that in the weak coupling limit, the $\nabla^4\cR^4$ coupling admits tree-level, one-loop and two-loop corrections, up to exponentially suppressed instanton corrections:
\be 
\label{d4r4weak}
\mathcal{E}_{(1,0)}^{(D),{\rm pert}} = g_{D}^{\;  \frac{2d+4}{d-8}} 
\left( \frac{\zeta(5)}{g_D^2} + \cE_{(1,0)}^{(d,1)} + g_D^2\, \cE_{(1,0)}^{(d,2)}  \right)
\ee
where
\bea
\label{dfourrfournew}
\cE_{(1,0)} ^{(d,1)}  & =& 
4\pi\xi(4) \, \int_{\cF_1} \de \mu_1(\rho) \, \Gamma_{d,d,1}(\rho) \, \cE(2,\rho)\ ,
\\
\label{dfourrfour2}
\cE_{(1,0)} ^{(d,2)}  &=& 4\pi \, \int_{\cF_2}  \de \mu_2(\Omega) \, \Gamma_{d,d,2} ( \Omega)\ .
\eea
Here, $\cE(s,\rho)=\tfrac{1}{2}\sum_{(c,d)=1} \frac{\rho_2^s}{|c\rho+d|^{2s}}$ 
the non-holomorphic Eisenstein series of $SL(2,\IZ)$, 
$\de\mu_h(\Omega)$ is the invariant measure on the Siegel upper-half plane of degree $h$,
normalized as in \cite{Florakis:2016boz},\footnote{This differs by 
a factor $2^{-h(h+1)/2}$ from the normalization used in \cite{Pioline:2015yea}.} $\Gamma_{d,d,h}(\Omega)$ is the genus-$h$ Siegel--Narain partition function,
\be
\Gamma_{d,d,h}(\Omega) = |\Omega_2|^{\frac{d}{2}}\, \sum_{Q\in(\IZ^{d,d})^{\otimes h}}
e^{\I\pi \Omega_{ij} p_L(Q_i) p_L(Q_j) - \I\pi \bar\Omega_{ij} p_R(Q_i) p_R(Q_j)} \ .
\ee
We do not display the dependence of $\Gamma_{d,d,h}(\Omega)$ on the torus moduli,
which enters through the projections $p_L(Q)$ and $p_R(Q)$, see below \eqref{ZtopLR}.
The decompactification limit of these modular integrals is computed in Appendix A using
the orbit method. For the one-loop amplitude, we find (see \eqref{Idsdec})
\bea
\label{d4r41loopR}
\cE_{(1,0)} ^{(d,1)} & = &R_s\, \cE_{(1,0)} ^{(d-1,1)} + 4\zeta(3) \xi(d-4) R_s^{d-4}+ 8\pi \xi(4) \xi(d+2) R_s^{d+2} 
\\
&&+ 16\pi \xi(4)\, R_s^{\frac{d+3}{2}} 
\sum_{\substack{Q\in \IZ_*^{d-1,d-1} \\(Q,Q) = 0}} \sigma_{d+1}(Q) 
\frac{K_{\frac{d+1}{2}}\left(2\pi R_s \sqrt{ p_L^2+p_R^2}\right)}{|p_L^2+p_R^2|^{\frac{d+1}{4}}} 
e^{2\pi\I( Q , a)} 
\nn\\
&&+ 16\pi \xi(3)\, R_s^{\frac{d-3}{2}} 
\sum_{\substack{Q\in \IZ_*^{d-1,d-1} \\(Q,Q) = 0}} \sigma_{d-5}(Q) 
\frac{K_{\frac{d-5}{2}}\left(2\pi R_s \sqrt{ p_L^2+p_R^2}\right)}{|p_L^2+p_R^2|^{\frac{d-5}{4}}} 
e^{2\pi\I( Q , a)} 
\nn\\
&&+ 16\pi R_s^{d-1} \sum_{\substack{Q\in \IZ_*^{d-1,d-1} \\ N=\tfrac12(Q,Q) \ne 0}} \sum_{n|Q} n^{d+1} \sigma_{3}(\tfrac{|N|}{n^2})  
\frac{B_{\frac{d-2}{2},\frac{3}{2}}\left(R_s^2 (p_L^2+p_R^2),R_s^2 |N|\right)}
{|N|^{3/2}} e^{2\pi \I(Q , a) }
\nn
\eea
where we denoted by $p_L\equiv p_L(Q),\,  p_R\equiv p_R(Q)$ the projections of $Q$ on the positive and negative 
$d-1$-dimensional planes in $\IR^{d-1,d-1}$ defining the moduli of $T^{d-1}$, by $a\in \IR^{2n-2}$ the
off-diagonal components of the metric and $B$-field on $S^1\times T^{d-1}$, $\sigma_s(Q)=\sigma_s({\rm gcd} Q)$ the divisor function,
and $B_{s,\nu}$ the integral 
\be
\begin{split}
\label{defBsnu}
B_{s,\nu}(x,y) = &  \int_0^{\infty} \frac{\de t}{t^{1+s}} e^{-\pi t -\pi x/t}\, K_\nu(2\pi y/t)\ ,\qquad x,y>0\\
=& \sum_{k=0}^{\infty} \frac{\Gamma(\nu+k+\tfrac12)\, K_{s-k-\frac12}\left(2\pi\sqrt{x+2y}\right)}
{k!\,\Gamma(\nu-k+\tfrac12)\,
(4\pi)^k\, y^{k+\frac12}\, (x+2y)^{\frac{2s-2k-1}{4}}}\,
\ .
\end{split}
\ee
It is worth noting that for $\nu\in \tfrac12+\IZ$, $B_{s,\nu}(x,y)$ collapses to  a finite linear combination
of ordinary modified Bessel functions, in particular for $\nu=\tfrac32$, the value relevant here, 
\be
B_{s,\frac32}(x,y) = (x+2y)^{\frac{1-2s}{4}} y^{-\frac12}\, K_{s-\frac12}\left(2\pi\sqrt{x+2y}\right)
+  \frac{1}{2\pi}(x+2y)^{\frac{3-2s}{4}} y^{-\frac32}\, K_{s-\frac32}\left(2\pi\sqrt{x+2y}\right)\ .
\ee
For conciseness however we shall express the result in terms of $B_{s,\frac32}(x,y)$.

For the two-loop contribution, we get instead (see \eqref{Idhdec})
\bea
\label{d4r42loopR}
\cE_{(1,0)} ^{(d,2)} &=& R_s^2\, \cE_{(1,0)} ^{(d-1,2)} + 2\xi(d-4) \, R_s^{d-3}\, \cE_{(1,0)} ^{(d-1,1)}
\\
&&+16\pi\,R_s^{\frac{d-1}{2}} \sum_{\substack{Q\in \IZ_*^{d-1,d-1} \\ (Q,Q) = 0}}  
\frac{\sigma_{d-5}(Q) }{(\mbox{gcd } Q)^{-1}}
 \cE^{D_{d-2}}_{\Lambda_{d-2}}(v_Q) \, 
 \frac{K_{\frac{d-5}{2}}\left(2\pi R_s \sqrt{p_L^2+p_R^2}\right)}
 {|p_L^2+p_R^2|^{\frac{d-3}{4}}} e^{2\pi \I( Q , a) } \nn
\eea
where $v_Q$ is the component of $v$ in $O(d-2,d-2)$, the Levi part of the stabilizer of the null vector $Q$ inside $O(d-1,d-1)$. It is worth noting that the one-loop term contains both Fourier coefficients supported on null vectors (originating from the constant terms in $\cE(2,\rho)$) and Fourier coefficients supported on generic vectors (originating from the non-zero Fourier coefficients in $\cE(2,\rho))$.
In contrast, the Fourier coefficients of the two-loop term are supported on null vectors only.

We now substitute the large radius expansions \eqref{d4r41loopR} and \eqref{d4r42loopR} in
the weak coupling expansion \eqref{d4r4weak}, and express the radius $R_s$ of the circle in string units in terms of the radius $R$ in Planck units in dimension $D+1$, and the string coupling $g_D$ in dimension $D$ by its counterpart  in dimension $D+1$ using 
\be
\label{Rtog1}
R=g_{D+1}^{\frac{2}{d-9}}\ ,\qquad g_{D+1}=(g_D \sqrt{R})^{\frac{9-d}{8-d}}\ .
\ee
In addition, we rewrite $p^2_L(Q)$ and $p^2_R(Q)$ in terms of the invariants $|Z(\Gamma)|^2$
and $\Delta(\Gamma)$ specialized to $\Gamma=(Q,0,0)\in \IZ^{6,6}\oplus \IZ^{32}\oplus \IZ^{6,6}$),
\be 
|Z(\Gamma)|^2 =g_{D+1}^{\; \frac{4}{9-d}}  \scal{Êp_L^2(Q) +  p_R^2(Q)} \ , \qquad 
\Delta(\Gamma) =\frac14 g_{D+1}^{\; \frac{8}{9-d}}  \left( p_L^2(Q) -  p_R^2(Q)\right)^2\ . 
\ee
Moreover, we observe that the  constraint $(Q,Q)=0$ is the specialization of the 
1/2-BPS constraint $\Gamma\times\Gamma=0$ for such $\Gamma$.
Finally, in the terms multiplying $K_{\frac{d-5}{2}}\left(2\pi R_s \sqrt{p_L^2+p_R^2}\right)$ in
\eqref{d4r41loopR} and \eqref{d4r42loopR}, we express $g_{D+1}$ in terms of a charge-dependent 
coupling 
\be
\label{Rtog2}
g_{D+2,\Gamma}= g_{D+1}^{\frac{d-10}{d-9}} \left( \frac{{\rm gcd}(\Gamma)}{|Z(\Gamma)|}\right)^{1/2}\ ,
\ee
which has the property of being invariant under $O(d-2,d-2)$ transformations stabilizing the 
vector $\Gamma=(Q,0,0)$.
The result of this procedure produces
\bea 
\label{d4r4large}
\mathcal{E}_{(1,0)}^{(D),{\rm pert}}  &&= 8 \pi R^{\frac{10}{8-d}} \Biggl( 
g_{D+1}^{\frac{2d+2}{d-9}} 
\left[\frac{\xi(2) \xi(5)}{g_{D+1}^2} Ê+ \xi(4) \xi(d+1)\, \cE^{D_{d-1}}_{\frac{d+1}{2} \Lambda_1}
+ \xi(2) \xi(4) g_{D+1}^{\;  2 }\, \cE^{D_{d-1}}_{2 \Lambda_{d-1}}\right]
\nn\\
+&&
 \xi(d-4)  R^{d-5} g_{D+1}^{\frac{2d-6}{d-9}} 
\left[Ê\frac{\xi(3)}{g_{D+1}^{2}} + \xi(2)\, \cE^{D_{d-1}}_{\Lambda_{d-1}}\right]Ê+ 
\xi(4) \xi(d+2) R^{d+1} \nn \\
+&&
 2 \xi(4) R^{\frac{d+1}{2}} \sum_{\substack{\Gamma\in \IZ_*^{d-1,d-1} \\  
 \Gamma \times \Gamma= 0}} \sigma_{d+1}(\Gamma)\, \frac{K_{\frac{d+1}{2}}(2\pi R | Z(\Gamma)|)}{|Z(\Gamma)|^{\frac{d+1}{2}}} e^{2\pi \I\langle\Gamma,a\rangle }  \nn \\ 
&&\hspace{-15mm}   +\quad 
 2 R^{\frac{d-5}{2}} \hspace{-5mm}\sum_{\substack{\Gamma\in \IZ_*^{d-1,d-1} \\ 
 \Gamma\times\Gamma= 0}}  \frac{\sigma_{d-5}(\Gamma) }{(\mbox{gcd } \Gamma)^{\frac{6}{d-10}} }
\left\{ \frac{\xi(3)  g_{D+2,\Gamma}^{\; -2}  + \xi(2)\,  \cE^{D_{d-2}}_{\Lambda_{d-2}}(v_\Gamma) }
{g_{D+2,\Gamma}^{\; \frac{2d-8}{10-d}}}  \right\}
\frac{K_{\frac{d-5}{2}}(2\pi R | Z(\Gamma)|)}{|Z(\Gamma)|^{\frac{d-5}{2}+\frac{6}{10-d}}} 
e^{2\pi \I\langle\Gamma,a\rangle }\nn \\
+&&
 2  R^{d-2} \sum_{\substack{\Gamma\in \IZ_*^{d-1,d-1} \\  \Gamma\times\Gamma \ne 0}} \sum_{n|\Gamma} n^{d+1} \sigma_{3}(\tfrac{\Gamma \times \Gamma}{n^2})  \frac{B_{\frac{d-2}{2},\frac{3}{2}}(R^2 | Z(\Gamma)|^2,R^2 \sqrt{\Delta(\Gamma)})}{\Delta(\Gamma)^{\frac{3}{4}}} 
 e^{2\pi \I\langle\Gamma,a\rangle } \Biggr)\
\eea
The terms on the first two lines are recognized as the weak coupling expansion of the $\nabla^4\cR^4$
coupling in dimension $D+1$ and of the expected massless threshold effects, proportional to $R^{d-5}$
and $R^{d+1}$, respectively \cite{Green:2010wi,Pioline:2015yea}. The third and fourth lines correspond to contributions from perturbative 1/2-BPS states winding around the circle, while the last line
corresponds to contributions from perturbative 1/4-BPS states. While these contributions are
invariant under the T-duality group $O(d-1,d-1)$ in dimension $D+1$, they are  not invariant under the full U-duality group $E_{d}(\IZ)$. However, invariance under $E_{d}(\IZ)$ can be restored 
by promoting the sum over $\Gamma\in \IZ_*^{d-1,d-1}$ to a sum over vectors in the full charge lattice 
$M^{E_{d}}_{\Lambda_d}=\IZ^{\dim(\Lambda_d)}$ in dimension $D+1$, subject to the 1/4-BPS constraint $I'_4(\Gamma)=0$,
and by replacing the term in braces on the fourth line by the Eisenstein series 
$\cE^{\scalebox{0.7}{$E_{d-1}$}}_{\scalebox{0.65}{$\frac32\Lambda_1$}}(g_\Gamma)$ for the stabilizer group $E_{\scalebox{0.6}{$d\, $-$1$($d\,$-$1$)}}$ of a 1/2-BPS charge vector --
which matches the said term at weak coupling. Thus, we conclude that the large radius expansion
of the exact $\nabla^4\cR^4$ coupling in $D$ dimensions is given by 
\bea 
\label{d4r4largestrong}
\mathcal{E}_{(1,0)}^{(D)} &&= R^{\frac{10}{8-d}} \Biggl( 
\mathcal{E}_{(1,0)}^{(D+1)}
+2 \xi(d-4)  R^{d-5} \, \mathcal{E}_{(0,0)}^{(D+1)}+ 
8\pi\xi(4)\, \xi(d+2) R^{d+1}  \\
+&&
 16\pi \xi(4) R^{\frac{d+1}{2}} \sum_{\substack{\Gamma\in M^{E_{d}}_{\Lambda_d* }\\  \Gamma\times\Gamma = 0}} \sigma_{d+1}(\Gamma)\, \frac{K_{\frac{d+1}{2}}(2\pi R | Z(\Gamma)|)}{|Z(\Gamma)|^{\frac{d+1}{2}}} e^{2\pi \I\langle\Gamma,a\rangle }  \nn \\ 
+&&
 16\pi \xi(3) R^{\frac{d-5}{2}} \sum_{\substack{\Gamma\in M^{E_{d}}_{\Lambda_d*} \\  \Gamma\times\Gamma= 0}}  \frac{\sigma_{d-5}(\Gamma)\, \cE^{E_{d-1}}_{\frac32\Lambda_1}(g_\Gamma)\ }{(\mbox{gcd } \Gamma)^{\frac{6}{d-10}} }\,
\frac{K_{\frac{d-5}{2}}(2\pi R | Z(\Gamma)|)}{|Z(\Gamma)|^{\frac{d-5}{2}+\frac{6}{10-d}}} 
e^{2\pi \I\langle\Gamma,a\rangle }\nn \\
+&&
 16\pi  R^{d-2}  \hspace{-3mm} \sum_{\substack{\Gamma\in M^{E_{d}}_{\Lambda_d*}\\  I'_4(\Gamma)=0, \Gamma\times\Gamma \ne 0}} \sum_{n|\Gamma} n^{d+1} \sigma_{3}(\tfrac{\Gamma \times \Gamma }{n^2})  \frac{B_{\frac{d-2}{2},\frac{3}{2}}(R^2 | Z(\Gamma)|^2,R^2 \sqrt{\Delta(\Gamma)})}{\Delta(\Gamma)^{\frac{3}{4}}} 
 e^{2\pi \I\langle\Gamma,a\rangle} \Biggr)+\dots
 \nn
\eea
where $a\in M_{\Lambda_d}^{E_d}\otimes\IR$ now includes the holonomies of all electromagnetic fields in dimension
$D+1$, and the dots denotes potential additional terms that are by construction exponentially suppressed in both the radius and the weak coupling expansion. We shall argue below that such terms are absent for $D>3$, and only contribute to non-Abelian Fourier coefficients for $D=3$. Eq. \eqref{d4r4largestrong} is
the main technical result of this paper.

Granting this claim for the moment, we see 
from the first line in  \eqref{d4r4largestrong} that in the limit $R\to\infty$, the $\nabla^4\cR^4$ coupling in dimension $D$ receives the expected power-like contributions in the radius, one proportional to the $\nabla^4 \cR^4$ coupling in dimension $D+1$, the two others being the first terms in an infinite series of powerlike contributions which sum up to the massless threshold contributions in dimension $D+1$. The second and third line in  \eqref{d4r4largestrong} correspond to  contributions from 1/2-BPS states
with charge $\Gamma$ and mass $\cM_{1/2}=|Z(\Gamma)|$, while the last line in  \eqref{d4r4largestrong} corresponds to contributions from 1/4-BPS states. Using the fact that 
\be
\label{Bsnulim}
B_{s,\nu}(x,y) \sim \frac{e^{-2\pi \sqrt{x+2y}}}{2 y^{1/2} (x+2y)^{s/2}}
\ee
for large values of $x,y$, we see that  a 1/4-BPS black hole with primitive charge $\Gamma$
contributes for large $R$ as
\be
\label{contrib14}
\frac{R^{d-3}\, \sigma_3(\Gamma\times\Gamma)}{\Delta(\Gamma)\, [\cM_{1/4}(\Gamma)]^{\frac{d-2}{2}}}\,  e^{-2\pi R \cM_{1/4}(\Gamma)+2\pi\I \langle\Gamma,a\rangle}\ ,
\ee
in particular, it is exponentially suppressed as $e^{-R \cM_{1/4}(\Gamma)}$, and proportional to the helicity supertrace $\Omega_{12}(\Gamma)$ in \eqref{om12}, as announced.\footnote{Here,  it may be worth noting that the divisor sum $\sigma_3$ in \eqref{om12} originates from the Fourier coefficients of $E_4$ in the helicity generating function \eqref{Omrho}, while $\sigma_3$ in \eqref{contrib14} originates from the Fourier coefficients of $\cE(2,\rho)$ in the one-loop amplitude \eqref{dfourrfournew}. Indeed, the two are related by $E_4 \propto D_2\, D_0\, \cE(2,\rho)$
where $D_w=\tfrac{\I}{\pi}(\partial_\rho-\frac{\I w}{2\rho_2})$ is the Maass raising operator. }
The fact
that the only exponential corrections come from 1/2-BPS and 1/4-BPS terms is in accordance with the expectation that the $\nabla^4\cR^4$ coupling is 1/4-BPS saturated.

For comparison, we state the  large radius expansion
of the exact $\cR^4$ coupling in $D$ dimensions (up to the non-Abelian Fourier coefficients in $D=3$), which can be extracted by exactly the same 
techniques (cf. \cite{Bossard:2015foa}):
\begin{multline} 
\label{r4largestrong}
\mathcal{E}_{(0,0)}^{(D)} = R^{\frac{6}{8-d}}  \Biggl( 
\mathcal{E}_{(0,0)}^{(D+1)}
+4\pi\, \xi(d-2)  R^{d-3}   \\
+4\pi R^{\frac{d-3}{2}} \sum_{\substack{\Gamma\in M^{E_{d}}_{\Lambda_d*}\\  \Gamma\times\Gamma = 0}} \sigma_{d-3}(\Gamma)\, \frac{K_{\frac{d-3}{2}}(2\pi R | Z(\Gamma)|)}{|Z(\Gamma)|^{\frac{d-3}{2}}} e^{2\pi \I\langle\Gamma,a\rangle }   \Biggl)
 \end{multline}
Here, we see that beyond the expected power-like threshold effects \cite{Green:2010wi}, the only exponentially
suppressed corrections come from 1/2-BPS states, in accordance with the expectation that the 
$\cR^4$ coupling is 1/2-BPS saturated \cite{Pioline:2010kb,Green:2011vz}.

Let us now make some comments about the above results. First, the fact that the (Abelian) Fourier
coefficients of the $\cR^4$ and $\nabla^4\cR^4$ couplings with respect to the maximal parabolic
subgroup $P_d$ relevant for the circle decompactification limit have support on 1/2-BPS and 
1/4-BPS charges, respectively, is a characteristic property of the automorphic forms 
attached to the minimal and next-to-minimal nilpotent orbits, respectively \cite{Green:2011vz}. 
This is also in agreement with the observation made at the beginning of Section 3, that the 
Gelfand--Kirillov dimension of these automorphic forms is equal to the dimension of the
homogeneous cones $\cC_{1/2}$ and $\cC_{1/4}$ inside the charge lattice $\Lambda$ (plus one in
$D=3$, due to the occurrence of Taub-NUT instantons). 

Second, in the case of the $\nabla^4\cR^4$ coupling, 1/2-BPS states contribute two types
of terms, corresponding to the second and third line of \eqref{d4r4largestrong}. The first
type is the product of a Bessel function of order $\tfrac{d+1}{2}$ times a divisor 
function of $\Gamma$, similar to the one seen in \eqref{r4largestrong}
for the $\cR^4$ coupling. The second type involves the product of  a Bessel function of order 
$\tfrac{d-5}{2}$
times an automorphic form under $E_{\scalebox{0.6}{$d\, $-$1$($d\,$-$1$)}}$, which is the Levi part of the stabilizer
of a 1/2-BPS charge vector such that $\Gamma\times\Gamma=0$. From the weak coupling
expansion, it appears that this automorphic form is the minimal theta series 
$\cE^{\scalebox{0.7}{$E_{d-1}$}}_{\scalebox{0.65}{$\frac32\Lambda_1$}}$, but one may wonder if a different automorphic form $\cE'$ under $E_{\scalebox{0.6}{$d\, $-$1$($d\,$-$1$)}}$ could appear,
which may not be distinguishable from the minimal theta series at weak coupling. 
A basic constraint comes from
requiring that the sum of the Gelfand--Kirillov dimension of $\cE'$ and of the dimension
of the cone $\cC_{1/2}$ of charges $\Gamma$ must not exceed the Gelfand--Kirillov dimension
of $\cE_{\scalebox{0.6}{$(1,0)$}}^{\scalebox{0.6}{$(D)$}}$. One checks from Table 1 that the only 
possibility is to have an automorphic form in the same representation as the Eisenstein series $\cE^{\scalebox{0.7}{$E_{d-1}$}}_{\scalebox{0.7}{$s\Lambda_1$}}$, but an arbitrary value of the parameter $s$ is in principle allowed by dimension counting. However, $s\ne \frac{3}{2}$ is excluded by the tensorial differential equations that $\cE_{\scalebox{0.6}{$(1,0)$}}^{\scalebox{0.6}{$(D)$}}$ must satisfy \cite{Bossard:2014lra,Bossard:2014aea}, as we show in Appendix \ref{DiffEqE8} for $D=3$. More generally, one would expect $\cE'$ to be an automorphic form for the full parabolic subgroup 
$P_d \subset E_{d(d)}$ 
stabilizing $\Gamma$, not only under its Levi subgroup $E_{\scalebox{0.6}{$d\, $-$1$($d\,$-$1$)}}$. The same argument relying on the Gelfand--Kirillov dimension implies that the sum of the  dimensions of the Fourier support of $\cE'$ on the unipotent radical $U_d$ of $P_d$ and the dimension of the  cone $\cC_{1/2}$ should not exceed the Gelfand--Kirillov dimension of the next-to-minimal representation. The minimal dimension for this Fourier support is the dimension of the cone $\cC_{1/2}$ in $D+2$ dimension. For $D>3$, it is easily checked again from Table 1 that the Fourier coefficient cannot depend on $U_d$, whereas for $D=3$ this  is in principle allowed by dimension counting: $17+18=45<46$. In $D=3$ one could therefore have an additional contribution to $\cE'$ which Fourier coefficients associated to 1/2-BPS charges $q$ on the unipotent radical $U_7$ would be a function $\cF$ of the $E_{6(6)}$ invariant mass $|v(q)|$, $|Z(\Gamma)|$ and $R$. We prove in Appendix \ref{DiffEqE8} that there is no solution of this type to the third order differential equation implied by supersymmetry \cite{Bossard:2014lra}. Moreover, we check that the Abelian Fourier coefficients for a 1/2-BPS charge $\Gamma$ that we obtain in this section are the unique solutions to this third order differential equation with the required asymptotic behaviour at large radius. 

We conclude therefore that the constraints coming from the next-to-minimal $E_{d+1}$ representation imply that the weak coupling expansion is sufficient to determine unambiguously the complete Fourier expansion of the function $\cE_{\scalebox{0.6}{$(1,0)$}}^{\scalebox{0.6}{$(D)$}}$ for $D>3$, and its Abelian component for $D=3$, so that the dots in \eqref{d4r4largestrong} only include the non-Abelian Fourier coefficients in $D=3$ and vanish for $D>3$.

\section{Fourier expansion from constrained lattice sum \label{sec_latsum}}

In this section, we shall verify the large radius expansion \eqref{d4r4largestrong}
of the next-to-minimal theta series by a direct analysis of the maximal parabolic Eisenstein series $\cE^{E_{d+1}}_{s\Lambda_{d+1}}$ represented as a constrained Epstein series  \cite{Obers:1999um}
\be
\cE^{E_{d+1}}_{s\Lambda_{d+1}}   = \frac{1}{2\zeta(2s)} \sum_{\substack{ {\cal Q}\in M^{E_{d+1}}_{\Lambda_{d+1}*} \\ 
\cQ \times \cQ=0}} | \mathcal{V}({\cal Q})|^{-2s} \ ,  \label{LEpstein} 
\ee
where $M^{E_{d+1}}_{\Lambda_{d+1}}$ denotes a suitable $\IZ$-module of $E_{d+1}(\IZ)$
with highest weight $\Lambda_{d+1}$, $| \mathcal{V}({\cal Q})|^2$ is the $E_{d+1}$-invariant
norm on this module,  and $\cQ\times \cQ=0$ is a quadratic constraint
which ensures that all symmetric powers $\cQ^{\otimes_s k}$ have support
on the $\IZ$-module with highest weight $k\Lambda_{d+1}$. The identity
\eqref{LEpstein} depends on the fact that the non-zero primitive vectors $\cQ$ such that $\cQ\times
\cQ=0$ lie in a single orbit of $E_{d+1}(\IZ)$ with stabilizer $P_{d+1}\cap E_{d+1}(\IZ)$,
where $P_{d+1}$ is the parabolic subgroup associated to the simple root $\alpha_{d+1}$. This was proved in \cite{krutelevich2007jordan} for $E_7$ and $E_6$ and we shall prove it for $E_8$ in this section. The large radius expansion of the sum \eqref{LEpstein} can be
analyzed by standard Poisson resummation techniques, at least for  classes of 
vectors $\mathcal{Q}$ which have support on the three highest grade eigenspaces with respect
to $P_{d+1}$ (for $d\le 5$, all vectors with $\cQ\times\cQ=0$ are in that class since the representation decomposes in three components). As we shall see, the contribution from these classes of vectors at $s=\tfrac{d-2}{2}$ 
and $s=\frac{d+2}{2}$ reproduces \eqref{r4largestrong} and \eqref{d4r4largestrong}, respectively, 
suggesting that the contributions of the remaining vectors for $d\geq 6$ vanish at the special values of $s$ above, or (for $d=7$) contribute only to non-Abelian Fourier coefficients. We directly attack the most involved case $d=7$, and then comment on simplifications which take place for $d=6$ and $d=5$.

\subsection{$E_8$ lattice sum}

For $d=7$, the highest weight $\Lambda_{d+1}$ corresponds to the adjoint representation ${\bf 248}$
of $E_8$. As explained in Appendix \ref{sec_e8lat}, the Lie algebra $\mathfrak{e}_{8(8)}$ contains
a lattice ${M^{E_{8}}_{\Lambda_8}}$ which is invariant under the Chevalley group $E_8(\IZ)$. Its structure is most easily described using the 5-grading
\be 
\mathfrak{e}_{8(8)} \cong {\bf 1}^{\ord{-2}} \oplus {\bf 56}^{\ord{-1}} \oplus \scal{ \mathfrak{gl}_1 \oplus \mathfrak{e}_{7(7)} }^\ord{0}  \oplus {\bf 56}^{\ord{1}} \oplus {\bf 1}^{\ord{2}}  \ . \label{E8E7}  
\ee
Decomposing $\cQ=(n,\Upsilon,\ell+Q,\Gamma,m)$ along \eqref{E8E7}, the  vectors $\cQ\in M^{E_{8}}_{\Lambda_8}$ are those for which $n,m$ are integers, $\Upsilon,\, \Gamma$ 
are elements of $\IZ^{56}$  and $Q$ in $\mathfrak{e}_{7(7)}$, $\ell\in \IZ/2$ such that $Q+\ell $ 
acts on $\IZ^{56}$ by multiplication with  an integral matrix. 
The invariant norm is given by
\bea
\label{e8norm}
 | \mathcal{V}(\mathcal{Q})|^2 &=& R^{-4}Ê\scal{Ê m +  \langle a,\Gamma+b \Upsilon \rangle + 2 b \ell + b^2 n +\tfrac{1}{2}Ê \langle a , Q\cdot a \rangle + \tfrac{1}{4}Ê\langle \Upsilon , I_4^\prime(a) \rangle + \tfrac{1}{4} n I_4(a)}^2 \nn  \\
&& + R^{-2} \bigl| Z\scal{Ê \Gamma + Q\cdot a + \ell a + \tfrac{1}{8} I_4^\prime(a,a,\Upsilon)  +\tfrac{1}{2} a  \langle a , \Upsilon \rangle+\tfrac{1}{4}n I_4^\prime(a)  + b ( \Upsilon + a n ) } \bigr|^2 \nn \\
&& +\bigl| V\scal{ÊQ + 2 a \times \Upsilon +a \times a n }\bigr|^2 + \scal{\ell + \tfrac{1}{2}Ê\langle a , \Upsilon \rangle + b n }^2 \nn \\
&& +R^2\bigl| Z(\Upsilon + a n)\bigr|^2 + R^4 n^2 \ , 
\eea
where $R\in\IR^+,a\in\IR^{56},b\in\IR$ parametrize the directions $\mathfrak{gl}_1^\ord{0}  \oplus {\bf 56}^{\ord{1}} \oplus {\bf 1}^{\ord{2}}$ in $E_{8(8)}/Spin(16)$, and $|V(Q)|$ and $|Z(\Gamma)|, |Z(\Upsilon)|$
are the $E_7$-invariant norms on $\mathfrak{e}_{7(7)}$ and $\IZ^{56}$, respectively. The  
condition $\cQ \times \cQ=0$, corresponding to the vanishing of the projection of the tensor product  ${\bf 248}\otimes_s {\bf 248}={\bf 1}\oplus {\bf 3875}\oplus {\bf 27000}$ on the ${\bf 3875}$ component,
automatically implies that ${\bf 248}^{\otimes_s k}$ has support on the highest-weight module $k\Lambda_8$ for all $k\geq 1$.  Decomposing the tensor product ${\bf 248}\otimes_s {\bf 248}$ as in \eqref{E8E7},\footnote{Recall
that ${\bf 133}\otimes_s{\bf 133}={\bf 1}\oplus {\bf 1539} \oplus {\bf 7371}$, ${\bf 133}\otimes{\bf 56} = {\bf 56}\oplus {\bf 912} \oplus {\bf 6480}$, ${\bf 56}\otimes_s {\bf 56}={\bf 133}\oplus {\bf 1463}$,
${\bf 56}^{\otimes 3}={\bf 56}\oplus {\bf 6480}\oplus {\bf 24320}$. 
}
\be 
{\bf 3875} \cong {\bf 133}^{\ord{-2}} \oplus ({\bf 912} \oplus {\bf 56})^{\ord{-1}} \oplus \scal{ {\bf 1539 } \oplus {\bf 133} \oplus {\bf 1} }^\ord{0}  \oplus ({\bf 912} \oplus {\bf 56})^{\ord{1}} \oplus {\bf 133}^{\ord{2}}  \ , \label{3875E7}  
\ee
we find that $\cQ \times \cQ=0$ amounts to the following conditions,  corresponding to the components of grade $-2$ through $2$ in \eqref{3875E7}, 
\bea 
i)\quad \Upsilon \times \Upsilon &=& n Q \ , \nn\\
ii)\quad \tfrac{1}{3}ÊQ \cdot \Upsilon &=& n \Gamma - \ell \Upsilon \ , \qquad 2\Upsilon \times ( Q\cdot J) + \tfrac{2}{3} J \times ( Q\cdot \Upsilon) = \langle J , \Upsilon\rangle \, Q   \ , \nn \\
iii)\quad Q^2 \cdot J &=&   (3 \ell^2 -3  m n + \tfrac{1}{2}Ê\langle \Upsilon , \Gamma \rangle ) J + 2 \Upsilon \langle \Gamma , J\rangle -   2 \Gamma \langle \Upsilon , J\rangle \ , \nn \quad \Upsilon \times \Gamma = \ell Q\ ,  \\
iv)\quad \tfrac{1}{3}Ê Q \cdot \Gamma &=& \ell \Gamma - m  \Upsilon \ , \qquad 2\Gamma \times ( Q\cdot J) + \tfrac{2}{3} J \times ( Q\cdot \Gamma) = \langle J , \Gamma \rangle \, Q \ , \nn \\
v)\quad\Gamma \times \Gamma &=& m Q \ .
\label{3875explicit}  
 \eea
Here,  $J$ is an arbitrary element in $\IZ^{56}$, which we use to enforce the vanishing of the components ${\bf 912}^{(\pm 1)}$ and $({\bf 1}\oplus {\bf 1539})^{(0)}$. 
The equality \eqref{LEpstein} was argued in \cite{Bossard:2015foa} by comparing the infinitesimal character and checking the normalisation by comparison with the Langlands constant term formula. Here we shall prove directly that all the elements ${\cal Q}\in M^{E_{8}}_{\Lambda_8*}$ 
such that $\cQ \times \cQ=0$ lie in the same orbit as 
$\cQ=(0,0,0+0,0,m)$ for a suitable integer $m\in \IZ_*$,
hence proving \eqref{LEpstein}, 
\be  
 \label{LEpsteinE8}
\sum_{\substack{ {\cal Q}\in { M^{E_{8}}_{\Lambda_8*}}\\ 
\cQ \times \cQ=0}} | \mathcal{V}({\cal Q})|^{-2s}  
= \sum_{m\in \IZ_*} \sum_{\gamma \in P_8 \backslash E_8(\IZ)}  
|m|^{-2s}ÊR^{4s}|_\gamma = 2 \zeta(2s) \cE^{E_8}_{s\Lambda_8} \ . 
\ee
For this purpose, we shall analyze the various branches a)-f) of solutions to \eqref{3875explicit}. 
\begin{itemize}
\item The generic branch a) corresponds to $n\ne 0$. It follows  from i) and the first equation in ii)
that $Q$ and $\Gamma$ are given by
\be 
Q = \frac{1}{n} \Upsilon \times \Upsilon \ , \qquad \Gamma = \frac{1}{4n^2} I_4^\prime(\Upsilon) + \frac{\ell}{n} \Upsilon \ . 
\ee
Because the symmetric product ${\bf 56}^{\otimes_s 3}$ does not include the ${\bf 912}$, the second equation in ii) is automatically satisfied. Using iii) and the algebraic identities \eqref{I4ident} one deduces the value of $m$,
\be 
m =   \frac{I_4(\Upsilon)}{4n^3} + \frac{\ell^2}{n}  \ .  
\ee 
One then checks that the remaining constraints are automatically satisfied by this solution. Ignoring quantization conditions, we see that the cone $\cQ\times \cQ=0$  inside $\mathfrak{e}_{8(8)}$ is parametrized by the 58 variables $\Upsilon, \ell, n$, in agreement with  
the dimension of the minimal nilpotent adjoint orbit of $E_{8(8)}$. The elements for which $n$ divides both $\Upsilon$ and $\ell$ are simply generated from the special solution $\cQ=(n,0,0,0,0)$ by acting with an  element of the unipotent subgroup \eqref{Taction}. One can prove that all the solutions are also in the $E_8(\IZ)$ orbit of $\cQ=(n,0,0,0,0)$ using the same strategy as in \cite{krutelevich2007jordan}. Any charge vector $\Upsilon$ can be brought  by a suitable $E_7(\IZ)$ transformation into the canonical form $\Upsilon = ( q_0 , {\rm diag}(q_i),0,p^0) $ \cite{krutelevich2007jordan}. Using elements of the form \eqref{Taction} 
\be 
\Upsilon \rightarrow \Upsilon  + n a 
\ee
one can 
obtain a new representative of the orbit for which all entries $q_0, q_i, p^0$ lie in $[0,n-1]$. Using then \eqref{CTaction} with a rank one $\bar a$ such that the only non-vanishing component is the one conjugate to the smallest of $q_0, q_i, p^0$ with respect to the symplectic form ({\it i.e.} such that $\langle \bar a , (q_0,q,0,p^0)\rangle =c \, {\rm min}\{ q_0, q_i, p^0\}$), one can modify $n$ to 
\be n\rightarrow n + c \, {\rm min}\{ q_0, q_i, p^0\} \ee
and therefore get a new representative such that  $n$ belongs to  $[1,{\rm min}\{ q_0, q_i, p^0\} ]$. Iterating these two steps, the values $q_0, q_i, p^0$ decrease by a non-zero amount in each step, and therefore vanish after a finite number of steps. We have therefore proved that any solution with a non-vanishing $n$ is in the same $E_8(\IZ)$ orbit as a representative for which $\Upsilon=0$ and $n\ne 0$. In this case \eqref{3875explicit} simplifies drastically and one finds that $Q=\Gamma=0$ while
\be \ell^2 = m n \ . \label{NullVec}Ê
\ee
It remains to prove that a null vector of $SO(2,1)$ can be rotated through the action of $SL(2,\IZ)
\subset E_8(\IZ)$ to a canonical representative for which only $n$ is non-zero. Let us decompose $n$ and $m$ into products of primes, $
n = \prod_p p^{n_p}, m = \prod_p p^{m_p}$
Because $n m$ is a square, if $n_p$ is odd then $m_p$ is odd as well such that the corresponding prime $p$ divides both $n$ and $m$, and one can always write the solution to \eqref{NullVec} as 
$ m = r y^2,\ell = r x y,n  = r x^2$
where $(x,y)$ transforms with respect to $SL(2,\IZ)$ as a doublet.  One can therefore use $SL(2,\IZ)$ to set $y=0$ and one obtains the desired property.

For the generic solution with $n\neq 0$, the invariant norm \eqref{e8norm} simplifies to 
\bea  
| \mathcal{V}(\mathcal{Q})|^2 &=& R^{-4}Ê\scal{Ê \tfrac{1}{4n^3} I_4(\Upsilon + n a )+ \tfrac{1}{n} ( \ell  + \tfrac{1}{2}Ê\langle a , \Upsilon\rangle+ b n )^2  }^2  \\
&& + R^{-2} \bigl| Z\scal{Ê \tfrac{1}{4n^2} I^\prime_4(\Upsilon + n a )+ \tfrac{1}{n} (\ell  +\tfrac{1}{2} \langle a , \Upsilon\rangle+ b n  )( \Upsilon + a n ) } \bigr|^2 \nn \\
&& +\bigl| V\scal{Ê\tfrac{1}{n} ( \Upsilon + a n ) \times ( \Upsilon + a n ) }\bigr|^2 + \scal{\ell +\tfrac{1}{2} \langle a , \Upsilon \rangle + b n }^2 \nn \\
&& +R^2\bigl| Z(\Upsilon + a n)\bigr|^2 + R^4 n^2\ ,  \nn
\eea
which is manifestly invariant under the action of the unipotent discrete group \eqref{Taction},
\bea a&\rightarrow& a+ u\ , \qquad b\rightarrow b+\tfrac{1}{2}Ê \langle a,u\rangle + k \ , \nn\\
\Upsilon &\rightarrow& \Upsilon - n  u\ , \qquad \ell \rightarrow \ell - n  k-\tfrac{1}{2}Ê\langle u , \Upsilon\rangle     \ .
\eea
Note that the discrete shift preserves the property that $m, \Gamma, Q+\ell$ are integer valued for $u\in \IZ^{56}$ and $k\in \IZ/2$ such that $\tfrac{1}{4}I_4(u)+k^2\in \IZ$, according to the definition of the $E_8(\IZ)$ unipotent subgroup (see Appendix \ref{sec_e8lat}). 

\item The second branch b) corresponds to $n=0$ but $\Upsilon\neq 0, \ell\neq 0$.
In this case the  constraint i) in \eqref{3875explicit} implies that $\Upsilon$ is rank one, namely 
$\Upsilon \times \Upsilon = 0$. It is convenient to decompose the constraints according to the $E_{6(6)} \ltimes \IR^{27}$ subgroup of $E_{7(7)}$ that stabilizes $\Upsilon$, using
\bea 
\mathfrak{e}_{7(7)} &\cong&  {\bf 27}^{\ord{-2}} \oplus \scal{ \mathfrak{gl}_1 \oplus \mathfrak{e}_{6(6)}  }^\ord{0}  \oplus \overline{\bf 27}^{\ord{2}}   \ ,  \nn\\
{\bf 56}  &\cong& {\bf 1}^{\ord{-3}} \oplus  \overline{\bf 27}^{\ord{-1}}  \oplus  {\bf 27}^{\ord{1}} \oplus  {\bf 1}^{\ord{3}} \ , \nn \\
{\bf 912}  &\cong& {\bf 78}^{\ord{-3}} \oplus \scal{Ê{\bf 351} \oplus  \overline{\bf 27}} ^{\ord{-1}}  \oplus  \scal{Ê \overline{\bf 351} \oplus {\bf 27}} ^{\ord{1}} \oplus  {\bf 78}^{\ord{3}} \ . \label{E7E6}
 \eea
The eigenvalue equation 
\be Q\cdot \Upsilon = - 3 \ell \Upsilon \ee
implies that $Q\in ( \mathfrak{gl}_1 \oplus \mathfrak{e}_{6(6)} )^\ord{0} 
\oplus \overline{\bf 27}^\ord{2}$, and that its
component along $ \mathfrak{gl}_1^\ord{0}$ is $-\ell$ where $\ell\in\IZ/2$. Moreover, since ${\bf 912}$ contains ${\bf 78}^\ord{3}$, the component of $Q$ along $ \mathfrak{e}_{6(6)}^\ord{0}$  must vanish. Using iv) in \eqref{3875explicit} one finds that $\Upsilon$ and $Q$ are given by
\be
\Upsilon = ( 0,0,0,r), \quad 
Q  = ( 0, -\ell+0,q) \ , \label{SolQ} 
\ee
while $\Gamma$ is the sum of two rank-one vectors, 
 \be 
 \Gamma = \Gamma_0 + \frac{m}{2\ell} \Upsilon\ ,\qquad 
 \Gamma_0=
  - \frac{1}{r} \bigl(4 \ell^2 ,2 \ell q,  q \times q , -\tfrac{1}{2\ell} \det q \bigr),
 \ee 
 such that
 \be 
 Q =\tfrac{1}{\ell} Ê\Upsilon \times \Gamma_0 \ ,\quad  \langle \Upsilon,\Gamma_0 \rangle = 4 \ell^2\ .
 \ee
One can check that  all equations in \eqref{3875explicit} are satisfied. 
In order for each entry in $\Gamma$ to be an integer, we must require that 
\be 
r| 4 \ell^2 \ , \qquad r|2 \ell q \ , \qquad r | q\times q \ , \qquad 2\ell | ( mr + \tfrac{\det q}{r}) \ . 
\ee
This solution is clearly in the same $E_8(\IZ)$ orbit as the generic branch solution since the action of \eqref{CTaction} with $\bar a=0$ gives $n= - 2 \bar b \ell + {\bar b}^2 m $ which does not vanish for a generic $\bar b$ since $\ell\neq 0$.   
 
\item We now consider the branch c) $n=\ell=0$ and $\Upsilon \ne 0$, for which we obtain  
 \be
\Upsilon = ( 0,0,0,r) \ , \quad 
Q    = ( 0,0+0,q) \ , \quad  \Gamma =   \bigl(0 , 0, \tfrac{1}{r} q \times q ,s  \bigr) \ ,\quad 
 m=  \frac{\det q}{r^2}  \ .  
 \ee
Using the action of \eqref{CTaction} with $\bar b = 0$ and $\bar a= ( 1,0,0,0)$ one obtains $n=r$ such that this solution is in the same $E_8(\IZ)$ orbit as the generic branch solution. 
  
\item We now turn to the case d) $n=\Upsilon=0, Q\neq 0$.  Then 
\be 
\forall J\ , Q^2\cdot J = 3 \ell^2 J \ , \qquad \ell Q = 0
\ee
implies that $\ell=0$ and $Q^2=0$. Since $Q$ belongs to the minimal adjoint orbit of 
$E_{7(7)}$, one can use a continuous $E_{7(7)}$ transformation  to it into the grade-two 
component of the $\mathfrak{so}(6,6)$ decomposition
\bea 
\mathfrak{e}_{7(7)} &\cong& {\bf 1}^{\ord{-2}} \oplus {\bf 32}_+^{\ord{-1}} \oplus \scal{ \mathfrak{gl}_1 \oplus \mathfrak{so}(6,6) }^\ord{0}  \oplus {\bf 32}_+^{\ord{1}} \oplus {\bf 1}^{\ord{2}} \ ,  \nn\\
{\bf 56}  &\cong& {\bf 12}^{\ord{-1}} \oplus {\bf 32}_-^{\ord{0}} \oplus {\bf 12}^{\ord{1}} \ ,\nn\\
{\bf 912}  &\cong& {\bf 32}_-^\ord{-2}\oplus ({\bf 220} \oplus {\bf 12})^{\ord{-1}}  \oplus {\bf 352}_+^{\ord{0}} \oplus ({\bf 220} \oplus {\bf 12})^{\ord{1}} \oplus{\bf 32}_-^\ord{2} \ . \label{E7D6}
 \eea 
The constraint  $Q\cdot\Gamma=0$ in the ${\bf 912}$ then 
implies that $\Gamma$ must lie in ${\bf 12}^{\ord{1}}$. The constraint v) in \eqref{3875explicit} 
implies that 
 \be 
 Q = ( 0,0,0,0,r) \ , \qquad \Gamma = ( 0,0,v) \ , \qquad (  v,v) = 2 r m \ . \label{VecSol}  
 \ee
If we assume that there is a non-trivial solution of this form such that $\Gamma \in \IZ^{56}$, then $Q = \tfrac{1}{m} \Gamma \times \Gamma$ and one can use $E_7(\IZ)$ to set $\Gamma$ in the canonical $E_{6(6)}$ form $\Gamma = ( 0,q,0,p^0)$ where $q$ is a diagonal matrix diag$(q_1,0,0)$ such that  $p^0|q_1$ \cite{krutelevich2007jordan}. Then  $Q = ( 0,0, \tfrac{p^0 q}{m}  )$, which is a canonical form with a $Spin(6,6,\IZ)\ltimes \IZ^{32+1}$ stabilizer. $ \langle a , Q\cdot a  \rangle$ defines indeed an even selfdual metric over $\IZ^{6,6}$ 
\be  
(  a^\ord{-1} , a^\ord{-1})  \equiv  \frac{1}{{\rm gcd} Q}Ê \langle a , Q\cdot a  \rangle = 2 \scal{Ê- u_1 v_1 + u_2 v_2 - \alpha \alpha^* }   \ , 
\ee
on the elements of the form (where $\alpha$ is an integral split octonion and $u_i,v_i$ are integers) 
\be 
a = \left(  u_1 , \begin{pmatrix} v_1 &\times&\times \\ \times &\times &\times   \\ \times &\times &\times \end{pmatrix},  \begin{pmatrix} \times  &\times &\times  \\ \times &u_2&\alpha   \\ \times &\alpha^*&v_2\end{pmatrix},\times \right)  \ . 
\ee
One can indeed prove in general that any element $Q$ of the minimal nilpotent orbit of $E_7$ over the integers can be rotated to a canonical form \eqref{VecSol}, such that there indeed always exist a non-trivial solution such that $\Gamma \in \IZ^{56}$, but we shall come back to this at the end of this discussion. 

The invariant norm \eqref{e8norm} evaluates to 
\bea
\label{e8normd}
| \mathcal{V}(\mathcal{Q})|^2 &= & 
R^{-4}Ê\scal{Ê m +  \langle a,\Gamma\rangle +\tfrac{1}{2}Ê \langle a , Q\cdot a \rangle }^2
+R^{-2}  \bigl| Z\scal{Ê \Gamma + Q\cdot a} \bigl|^2
+\bigl| V\scal{ÊQ} \bigl|^2\nn\\
&=& R^{-4}Ê\scal{Ê m +  (  a, v)  + \tfrac{r}{2}  (  a, a) }^2
+R^{-2} e^\phi \, g(v+a  n,v+an) + e^{2\phi} r^2 \; ,  \hspace{5mm} 
\eea
where in the second line, we denoted $a^\ord{-1}=a$ and introduced the notations
 \be 
 | V(Q)|^2 = e^{2\phi} r^2 \ , \qquad |Z(\Gamma)|^2 = e^{\phi} g(v,v) \ , 
 \ee
so that $e^\phi$ and $g(v,v)$ parametrize the moduli in $\IR^+\times SO(6,6)/( SO(6) \times SO(6))\subset E_{7(7)}/SU(8)$. It is worth noting that unlike branches $a),b),c)$, the norm
\eqref{e8normd} does not depend on the NUT axion $b$.

Using the action of \eqref{CTaction} with $\bar b = 0$ and $\bar a$ a vector of grade $-2$ of norm $2$ one obtains $n=r$ such that this solution is in the same $E_8(\IZ)$ orbit as the generic branch solution. 

\item We now turn to the branch  e) $n=\Upsilon = \ell = Q=0, \Gamma\neq 0$. 
In this case $\Gamma$ has rank one
($\Gamma \times \Gamma = 0 $) and the invariant bilinear form \eqref{e8norm} reduces 
 \be | \mathcal{V}(\mathcal{Q})|^2 = R^{-4} \scal{Ê m +  \langle a,\Gamma \rangle}^2  + R^{-2}Ê|Z(\Gamma)|^2 \ .  \ee 
 Using the action of \eqref{CTaction} with $\bar b = 0$ and $\bar a$ an element satisfying to $I_4(\bar a)=0$ and $\langle \Gamma , I_4^\prime(\bar a)\rangle \ne 0$ one obtains $n=-\langle \Gamma , I_4^\prime(\bar a)\rangle$ such that this solution is in the same $E_8(\IZ)$ orbit as the generic branch solution. To prove that such an element exists it is enough to consider the case $\Gamma = ( r,0,0,0)$ since all rank-one charge vectors $\Gamma$ are in the same $E_7(\IZ)$ orbit, and then choosing $\bar a = (0,0,p,0)$ one obtains $n = r \det p$. 
  
\item Finally, the last branch  f) is when $n=\Upsilon = \ell = Q= \Gamma = 0, m\neq 0$, in which case  $| \mathcal{V}(\mathcal{Q})|^2 = R^{-4} m^2$.  Using the action of \eqref{CTaction} with $\bar b = 1$ and $\bar a=0$  one obtains $n=m$ such that this solution is in the same $E_8(\IZ)$ orbit as the generic branch solution. 
\end{itemize}

To complete the proof that all elements of the minimal nilpotent orbit of $E_8$ in $ M^{E_{8}}_{\Lambda_8}$ are in the $E_8(\IZ)$ orbit of a canonical element of the form $(0,0,0,0,m)$ it remains to prove that all elements of the minimal nilpotent orbit of $E_7$ in the corresponding $ M^{E_{7}}_{\Lambda_1}$ are in the $E_7(\IZ)$ orbit of a canonical element of the form $(0,0,0,0,m)$ according to \eqref{VecSol}. To do this we note that the construction of the lattice $ M^{E_{8}}_{\Lambda_8}$ in Appendix \ref{sec_e8lat} and the chain of arguments developed in this section are valid for any group constructed in the same way starting from different split Jordan algebras as $E_{7(7)},\, E_{6(6)}$ and $SO(4+n,4+n)$ \cite{krutelevich2007jordan}. Applying the same construction to the adjoint lattice of $E_7$ one obtains that it remains to prove the equivalent result for $SO(6,6)$, and recursively for $SO(4,4)$ and finally $SL(2)$ for which it it true as we have proved below equation \eqref{NullVec}.\footnote{Recall that \bea \mathfrak{so}(6,6) &\cong& {\bf 1}^\ord{-2} \oplus ({\bf 2}\otimes {\bf 8})^\ord{-1} \oplus \scal{Ê\mathfrak{gl}_1\oplus \mathfrak{sl}_2 \oplus \mathfrak{so}(4,4)}^\ord{0}  \oplus  ({\bf 2}\otimes {\bf 8})^\ord{1} \oplus{\bf 1}^\ord{2} \nn\\
 \mathfrak{so}(4,4) &\cong& {\bf 1}^\ord{-2} \oplus ({\bf 2}\otimes {\bf 2}\otimes {\bf 2})^\ord{-1} \oplus \scal{Ê\mathfrak{gl}_1\oplus \mathfrak{sl}_2 \oplus \mathfrak{sl}_2\oplus \mathfrak{sl}_2}^\ord{0}  \oplus  ({\bf 2}\otimes {\bf 2}\otimes {\bf 2})^\ord{1} \oplus{\bf 1}^\ord{2} \nn
\eea and that the element of a minimal nilpotent orbit inside a semi-simple Levi component is the sum of elements of the minimal nilpotent orbits of the simple groups defining the Levi. In this case the nilpotency of the element of $\mathfrak{so}(6,6)$ in the ${\bf 32}_\pm$ and the ${\bf 12}$ implies separately  the element in $\mathfrak{sl}_2$ to be nilpotent in the fundamental and the element in $\mathfrak{so}(4,4)$ to be nilpotent in the thee 8-dimensional representations related by triality.}

 
We now decompose  the Epstein series \eqref{LEpsteinE8}  according to the six branches of solutions just described. 
The sum over elements supported on branch f) trivially gives 
\be 
\label{E8Eisf}
\cE^{E_8\ord{f} }_{s\Lambda_8} =  R^{4s} \ .  
\ee
The sum over elements on branch e) can be carried out by Poisson resummation over $m$, and 
gives
\bea 
\label{E8Eise}
\cE^{E_8\ord{e} }_{s\Lambda_8} &=&
\frac{R^2}{2\xi(2s)} \int_0^{\infty} \frac{\de t}{t^{s+\frac12}} \sum_{\tilde m\in\IZ} 
\sum_{\substack{\Gamma \in \IZ_*^{56} \\ \Gamma \times \Gamma = 0}} 
e^{-\pi R^4 \tilde m^2 t -\frac{\pi}{R^2 t} |Z(\Gamma)|^2+2\pi\I\tilde m \langle \Gamma,a\rangle}
\\
&&\hspace{-15mm}=
\frac{\Gamma(s-\tfrac12)\,R^{2s+1}}{2\xi(2s)\,\pi^{s-\frac12}}\, \sum_{\substack{ \Gamma\in\IZ^{56}_* \\ \Gamma\times\Gamma=0}}
|Z(\Gamma)|^{-(2s-1)}
+ \frac{2R^{\frac{6s+1}{2}} }{\xi(2s)} \hspace{-2mm}\sum_{\substack{\Gamma \in \IZ_*^{56} \\ \Gamma \times \Gamma = 0}}  \sigma_{2s-1}(\Gamma)  \frac{K_{s-\frac{1}{2}}( 2\pi R |Z(\Gamma)|)}{ |Z(\Gamma)|^{s-\frac{1}{2}}} e^{2\I\pi \langle \Gamma, a\rangle }\ .\nn
\eea
The first term, originating from $\tilde m=0$, is recognized as a maximal parabolic Poincar\'e series
for $P_7\backslash E_7(\IZ)$, leading to \cite{Bossard:2015foa}
\be 
\label{E8Eise2}
\cE^{E_8\ord{e} }_{s\Lambda_8} =  \frac{\xi(2s-1)}{\xi(2s)} R^{2s+1} \cE^{E_7}_{(s-\frac12)\Lambda_7} + \frac{2R^{\frac{6s+1}{2}} }{\xi(2s)} \hspace{-2mm}\sum_{\substack{\Gamma \in \IZ_*^{56} \\ \Gamma \times \Gamma = 0}}  \sigma_{2s-1}(\Gamma)  \frac{K_{s-\frac{1}{2}}( 2\pi R |Z(\Gamma)|)}{ |Z(\Gamma)|^{s-\frac{1}{2}}} e^{2\I\pi \langle \Gamma, a\rangle } 
\ .
\ee

We now turn to the contribution of the  elements 
$\cQ=(0,0,Q+\ell,\Gamma,m)$ supported on branch d). According to the discussion around \eqref{VecSol}, any solution on this branch is obtained from \eqref{VecSol} by an $E_7(\IZ)$ element,
and the stabilizer of \eqref{VecSol} is the parabolic subgroup $P_1$ with Levi subgroup 
$\IR^+\times SO(6,6)$. Changing $(v,r)$ to $(q,n)$ for convenience, 
  the contribution of branch d) can be written as a Poincar\'e sum 
\bea 
\label{E8Eisd0}
\cE^{E_8\ord{d}}_{s\Lambda_8}   &=& \frac{1}{\xi(2s)}
\sum_{\gamma \in P_1\backslash E_7(\IZ)} \gamma \biggl[ Ê\int_0^\infty \frac{\de t}{t^{1+s}}  \sum_{n>0}Ê\sum_{q\in \IZ^{6,6}} \sum_{m\in \IZ} \int_{-\frac{1}{2}}^{\frac{1}{2}} 
\de \rho_1 e^{ \I\pi  \rho_1( 2 mn-(q,q)) } \nn \\
&& \qquad \times e^{-\frac{\pi}{t} \scal{R^{-4} ( m+ (q,a) + \frac{(a,a)}{2} n )^2 + R^{-2} e^{\phi} g(q+a n,q+a n) + e^{2\phi} n^2} }\biggr] 
\eea
where we introduced a Lagrange multiplier $\rho_1$ for the constraint $(q,q)=2mn$.
After substituting $t = \frac{e^\phi}{R^2 \rho_2} $
one can rewrite this as an integral over the Poincar\'e upper half-plane strip,
\bea  
\label{E8Eisd1}
\cE^{E_8\ord{d}}_{s\Lambda_8}   &=& \frac{R^{2s}}{\xi(2s)}Ê 
\sum_{\gamma \in P_1\backslash E_7(\IZ)} \gamma \biggl[  e^{-s \phi} \int_{\mathcal{H}/\IZ} \frac{\de^2\rho}{\rho_2^{\; 2}} \rho_2^{s-\frac{5}{2}} \sum_{n>0}Ê\sum_{q\in \IZ^{6,6}} \sum_{m\in \IZ}   \nn \\
&& \hspace{-5mm}  \times \rho_2^{\; \frac{7}{2}} e^{-\pi \rho_2  \scal{R^{-2}e^{-\phi}  ( m+ (q,a) + \frac{(a,a)}{2} n )^2 +  g(q+a n,q+a n) + R^2 e^{\phi} n^2}+ \I\pi \rho_1 ( 2 mn-(q,q))  } \biggr] 
\eea
It is now convenient to rewrite the sum over strictly positive $n$ to one half the sum over all integral $n$ minus the case $n=0$, such that the  sum over $n\in\IZ$ is recognised as a standard Siegel--Narain theta series for a  lattice of signature $(7,7)$, while the second can be computed explicitly,
\bea  
\cE^{E_8\ord{d}}_{s\Lambda_8}   &=& \frac{R^{2s}}{2\xi(2s)}Ê 
\sum_{\gamma \in P_1\backslash E_7(\IZ)} \gamma \biggl[  e^{-s \phi}  \int_{\mathcal{H}/\IZ} \frac{\de^2\rho}{\rho_2^{\; 2}} \rho_2^{s-\frac{5}{2}}   \nn \\
&&  \times \rho_2^{\; \frac{7}{2}} Ê\sum_{q\in \IZ^{6,6}} \sum_{m,n\in \IZ} 
e^{-\pi \rho_2  \scal{R^{-2}e^{-\phi}  ( m+ (q,a) + \frac{(a,a)}{2} n )^2 +  g(q+a n,q+a n) + R^2 e^{\phi} n^2}+ \I\pi\rho_1  ( 2 mn-(q,q))  } \nn \\ 
& & \hspace{-5mm} -   e^{-s \phi}  \int_{\mathcal{H}/\IZ} \frac{\de^2\rho}{\rho_2^{\; 2}} \rho_2^{s-\frac{5}{2}} Ê\sum_{q\in \IZ^{6,6}} \sum_{m\in \IZ}   \rho_2^{\; \frac{7}{2}} e^{-\pi \rho_2  \scal{R^{-2}e^{-\phi}  ( m+ (q,a)  )^2 +  g(q,q) }- \I\pi \rho_1  (q,q)   } \biggr]\ .  \qquad 
\eea
Now the first term in the square bracket 
 can be folded into an integral  of the real analytic Eisenstein series of weight $s-\tfrac{5}{2}$ times the same Narain lattice sum
over the standard fundamental domain $\cF_1$ of $SL(2,\IZ)$: 
\bea  
&&\int_{\mathcal{F}_1} \frac{\de^2 \rho}{\rho_2^{\; 2}} \cE(s-\tfrac{5}{2},\rho) Ê\sum_{\substack{q\in \IZ^{6,6} \\ m,n\in \IZ}} \rho_2^{\; \frac{7}{2}} e^{-\pi \rho_2  \scal{\frac{ ( m+ ( q,a) + \frac{( a,a)}{2} n )^2}{ R^2 e^\phi}  +  g(q+a n,q+a n) + R^2 e^{\phi} n^2}+ \I\pi\rho_1  ( 2 mn-(q,q))  }
\nn \\ 
&&-  \int_{\mathcal{H}/\IZ} \frac{\de^2 \rho}{\rho_2^{\; 2}} \rho_2^{s-\frac{5}{2}} Ê\sum_{q\in \IZ^{6,6}} \sum_{m\in \IZ}   \rho_2^{\; \frac{7}{2}} e^{-\pi \rho_2  \scal{R^{-2}e^{-\phi}  ( m+ ( q,a)  )^2 +  g(q,q) }- \I\pi \rho_1 ( q,q)   } \nn 
\\
&=&e^{\frac{1}{2} \phi} R \hspace{-0.7mm} \int_{\mathcal{F}_1} \hspace{-0.7mm} \frac{\de^2 \rho}{\rho_2^{\; 2}} \cE(s-\tfrac{5}{2},\rho) Ê
\hspace{-1.5mm}\sum_{\substack{q\in \IZ^{6,6} \\ m,n\in \IZ}} \rho_2^{\; 3} e^{-\pi \rho_2  g(q+a n,q+a n)  -\pi R^2 e^{\phi} \frac{|\tilde m+ \rho n|^2}{\rho_2}   + \I\pi \tilde m ( 2 ( q,a) +( a,a)n ) -  \I\pi\rho_1  (q,q)  } \nn 
 \\ 
&&-  e^{\frac{1}{2} \phi} R 
\int_{\mathcal{H}/\IZ} \frac{\de^2 \rho}{\rho_2^{\; 2}} \rho_2^{s-\frac{5}{2}} Ê
 \sum_{q\in \IZ^{6,6}} \sum_{\tilde m\in \IZ}   \rho_2^{\; 3} e^{-\pi \rho_2 g(q,q) - \pi R^2 e^\phi\frac{\tilde m^2}{\rho_2}Ê+  2\I\pi \tilde m  ( q , a) - \I\pi \rho_1 ( q,q)   }
\eea
where in the second equality we performed a Poisson resummation over $m$.
Since $(\tilde m,n)$ transforms as a doublet under $SL(2,\IZ)$,
the integral over $\cF_1$ can be computed by 
applying the orbit method on $(\tilde m,n)$. For the trivial orbit  $(\tilde m,n)=(0,0)$,
the integral over $\cF_1$ can be 
again unfolded onto the strip by replacing the Eisenstein series by its seed, and the result cancels precisely the $\tilde m=0$ term in the second sum. For the non-trivial orbit one unfolds the modular domain back to the strip by restricting to $n=0$, obtaining in this way 
\bea  
\cE^{E_8\ord{d}}_{s\Lambda_8}   &=& \frac{R^{2s+1}}{\xi(2s)}Ê \sum_{\gamma \in P_1\backslash E_7(\IZ)} 
\gamma \biggl[   e^{-(s-\frac12) \phi}  \int_{\mathcal{H}/\IZ} \frac{\de^2 \rho}{\rho_2^{\; 2}} \scal{Ê\cE(s-\tfrac{5}{2},\rho) -  \rho_2^{s-\frac{5}{2}} Ê}  \hspace{20mm}  \nn \\
&& \hspace{30mm}  \times Ê\sum_{q\in \IZ^{6,6}} \sum_{\tilde m>0}   \rho_2^{\; 3} e^{-\frac{\pi}{ \rho_2}  R^2 e^{\phi} \tilde m^2 -\pi \rho_2  g(q,q) - \I\pi \rho_1 (q,q) + 2 \I\pi (\tilde m q , a)   } \biggr] \ .  \hspace{10mm}
\eea
One can then insert the Fourier expansion of $\cE(s-\tfrac{5}{2},\rho)$ \eqref{chowla},
\be \xi(2s-5)  \scal{Ê\cE(s-\tfrac{5}{2},\rho) -  \rho_2^{\; s-\frac{5}{2}} }  = 
\xi(2s-6) \rho_2^{\; \frac{7}{2} -s} \hspace{-1mm}+ 2  \hspace{-1mm}\sum_{N\in \IZ_*} \hspace{-1mm} \frac{\sigma_{2s-6}(|N|) }{|N|^{s-3}} \sqrt{ \rho_2}ÊK_{s-3}(2\pi |N| \rho_2)    e^{2\I\pi N \rho_1} 
\ee
to obtain
\bea  
\label{E8Eisd}
\xi(2s)\xi(2s-5) \cE^{E_8\ord{d}}_{s\Lambda_8}   &=& \xi(2s-6)\xi(2s-11)R^{12} Ê \sum_{\gamma \in P_1\backslash E_7(\IZ)}  \gamma\left[ e^{-2(s-3)\phi} \right] \\
&&\hspace{-40mm}  + 2 \xi(2s-6)   R^{s+\frac{13}{2}} \hspace{-5mm}Ê
\sum_{\gamma \in P_1\backslash E_7(\IZ)}\hspace{-2mm} \gamma  \Biggl[   e^{-2(s-3) \phi} \hspace{-2mm}\sum_{\substack{q\in \IZ^{6,6}_*\\ (q,q) = 0}}\sum_{m|q}  \Bigl( \frac{ \scalebox{0.8}{$\sqrt{Êe^\phi g(q,q)}$}}{m^2}\Bigr)^{s-\frac{11}{2}}Ê\hspace{-1mm}K_{s-\frac{11}{2}}
\left(2\pi R\scalebox{0.8}{$\sqrt{Êe^\phi g(q,q)}$}\right) e^{2\I\pi (q,a) }  \Biggr] \nn \\
&&\hspace{-40mm}  + 2  R^{2s+6} \hspace{-3mm}Ê\sum_{\gamma \in P_1\backslash E_7(\IZ)} 
\gamma  \Biggl[   e^{-(s-3) \phi} \sum_{\substack{q\in \IZ^{6,6}_*\\ (q,q) \ne  0}}\sum_{m|q}   m^{2s-1}
\left|Ê\frac{(q,q)}{2}\right|^{3-s} \sigma_{2s-6}(|\tfrac{(q,q)}{2m^2}|) \nn \\
&& \hspace{10mm} \times \int_0^{\infty} \frac{\de t}{t^{1+\frac{5}{2}}} 
e^{-\pi t - \frac{\pi}{t} R^2 e^\phi g(q,q)} K_{s-3}\left(\tfrac{2\pi}{t} R^2 e^\phi |\tfrac{(q,q)}{2}|\right)  
e^{2\I\pi (q,a) }  \Biggr]  \nn
\eea
where the fist term is the contribution of $q=0$, while for the last term we used the change of variable 
$\rho_2 = \frac{R^2 e^\phi m^2}{t}$ before reabsorbing the factor of $m$ in $q$ everywhere. We
shall now interpret the various contributions in terms of Langlands--Eisenstein series.

The first line is immediately recognized as $\cE^{E_7}_{(s-3)\Lambda_1}$.
 The  second line can be rewritten by noting that the sum over $\gamma\in P_1\backslash E_7(\IZ)$ together with the sum over null vectors $q\in\IZ_*^{6,6}$ can be reinterpreted as a sum over primitive $Q\in\IZ^{133}$ and (non-necessarily primitive) $\Gamma\in\IZ^{56}$ subject to the conditions
\be
\label{QGammacond}
Q^2 =0\ ,\quad \Gamma\times \Gamma=0\ ,\quad Q \, \Gamma|_{{\bf 912}\oplus {\bf 56}}=0\ . 
\ee
Indeed, these are the conditions  iv) and v) of \eqref{3875explicit} in the special
case $m=0$, which were solved in terms of the null vector $v\equiv q$ and $r\equiv n$
in \eqref{VecSol}, up to a element in $\gamma\in  E_7(\IZ)$. Now, we can 
solve the same conditions by instead using the decomposition \eqref{E7E6}, where 
$E_6$ is the Levi subgroup of the stabilizer of the rank one vector $\Gamma$. This means
that, up to an element $\gamma\in E_7(\IZ)$, the vector $\Gamma$ can be chosen
as $\Gamma = ( 0,0,0,\pm {\rm gcd} \Gamma)$. The constraint $Q \, \Gamma|_{{\bf 912}}=0$ requires
that $Q= (0,\ell+0,p)$, where $p\in \IZ^{27}$, while the constraint $Q^2=0$ requires $p\times p=0$ and $\ell = 0$.
Further decomposing $E_6$ into $SO(5,5)$, such a $p\in {\bf 10}^\ord{-2} \oplus {\bf 16}^\ord{1} \oplus {\bf 1}^\ord{4}$ can be further rotated into the highest component ${\bf 1}^\ord{4}$, invariant under
the parabolic subgroup $P_6$ of $E_6(\IZ)$, and the condition that $Q$ is primitive implies that this component is equal to $1$. Therefore, the sum over $Q,\Gamma$
subject to \eqref{QGammacond} can be recast into a sum over rank-one charge vectors
$\Gamma$ and a Poincar\'e sum over $\gamma\in P_6\backslash E_6(\IZ)$. 
Using the change of variable \footnote{According to the decomposition \eqref{E7E6} $ |Z(\Gamma)|^2 = e^{6\phi_7} ( {\rm gcd} \Gamma)^2 $ and $|V(Q)|^2 = e^{4\phi_7} |\vI(p)|^2 = e^{4\phi_7} e^{2\upsilon}|_\gamma$.} 
\be e^\phi g(q,q) = |Z(\Gamma)|^2 \ , \qquad e^{2\phi}  = e^{2\upsilon} \Bigl(\frac{|Z(\Gamma)|}{{\rm gcd} \Gamma} \Bigr)^\frac{4}{3} \ , \ee
where
$ e^{2\upsilon}$ is the character of the parabolic subgroup $P_1$ such that $|\vI(p)|^2 = e^{2\upsilon}|_\gamma$,
and we can rewrite the second line of \eqref{E8Eisd} as
\bea &&
 \sum_{\gamma \in P_1\backslash E_7(\IZ)} \gamma  \Biggl[   e^{-2(s-3) \phi} \sum_{\substack{q\in \IZ^{6,6}_*\\ (q,q) = 0}}\sum_{m|q}  \Bigl( \frac{ \scalebox{0.8}{$\sqrt{Êe^\phi g(q,q)}$}}{m^2}\Bigr)^{s-\frac{11}{2}}ÊK_{s-\frac{11}{2}}(2\pi R\scalebox{0.8}{$\sqrt{Êe^\phi g(q,q)}$}) e^{2\I\pi (q,a) }  \Biggr]  \nn \\
&=& \sum_{\substack{\Gamma \in \IZ^{56}_*\\ \Gamma \times \Gamma  = 0}} ({ \rm gcd } \Gamma)^{\frac{4}{3}(s-3)} \sigma_{11-2s}(\Gamma) \,\cE^{E_6}_{(s-3)\Lambda_1}(g_\Gamma) \,\frac{ K_{s-\frac{11}{2}}(2\pi R |Z(\Gamma)|)}{|Z(\Gamma)|^{\frac{2s+9}{6}} }   e^{2\I\pi \langle \Gamma ,a\rangle } \ .   
\eea

As for the third line in \eqref{E8Eisd}, we note that the sum over $\gamma\in P_1\backslash E_7(\IZ)$
and generic vectors $q$ in $\IZ_*^{6,6}$ can be recast as a sum over all rank 2 vectors 
$\Gamma\in\IZ^{56}_*$. Indeed, for any such vector, $\Gamma\times\Gamma$ 
belongs to the minimal nilpotent adjoint orbit of $E_{7(7)}$, and as such can be rotated through an element of $E_{7}(\IZ)$ to an integer representative of weight 2 in the decomposition \eqref{E7D6}. The vector $\Gamma$ then belongs to the component ${\bf 12}^\ord{1}$ in \eqref{E7D6}. Hence all 1/4-BPS charges are in the $E_{7}(\IZ)$ orbit of a generic vector.  

Putting these observations together and adding in \eqref{E8Eisf}, \eqref{E8Eise2}, we conclude that the complete sum over elements
in branches $def)$ gives
\bea  
\label{expE8s3}
\xi(2s)\,\xi(2s-5)\, \cE^{E_8\ord{def}}_{s\Lambda_8}   &=& 
\xi(2s)\,\xi(2s-5)\, R^{4s} + 
\xi(2s-1)\, \xi(2s-5)\, R^{2s+1} \cE^{E_7}_{(s-\frac12)\Lambda_7} 
\nn\\
&&+ 
\xi(2s-6)\, \xi(2s-11) \,R^{12} Ê \, 
\cE^{E_7}_{(s-3)\Lambda_1}   
\nn \\
&&
\hspace{-20mm} 
+ 2 \xi(2s-1)\,\xi(2s-5)\, R^{\frac{6s+1}{2}}  \hspace{-2mm}\sum_{\substack{\Gamma \in \IZ_*^{56} \\ \Gamma \times \Gamma = 0}}  \sigma_{2s-1}(\Gamma)  \frac{K_{s-\frac{1}{2}}( 2\pi R |Z(\Gamma)|)}{ |Z(\Gamma)|^{s-\frac{1}{2}}} e^{2\I\pi \langle \Gamma, a\rangle } 
 \nn \\
&&\hspace{-40mm}  + 2 \xi(2s-6)   R^{s+\frac{13}{2}} \sum_{\substack{\Gamma \in  \IZ_*^{56} \\ \Gamma \times \Gamma  = 0}} ({ \rm gcd } \Gamma)^{\frac{4}{3}(s-3)} 
\sigma_{11-2s}(\Gamma) \,
\cE^{E_6}_{(s-3)\Lambda_1}(g_\Gamma) \frac{ K_{s-\frac{11}{2}}(2\pi R |Z(\Gamma)|)}{|Z(\Gamma)|^{\frac{2s+9}{6}} }   e^{2\I\pi \langle \Gamma ,a\rangle } \nn \\
&&\hspace{-40mm}  + 2  R^{2s+6} \hspace{-6mm} \sum_{\substack{\Gamma \in   \IZ_*^{56} \\ 
\Gamma \times \Gamma  \ne  0,\, I_4^\prime(\Gamma)=0}} \hspace{-2mm} \sum_{m|\Gamma}   m^{2s-1} \sigma_{2s-6}(\tfrac{\Gamma \times \Gamma}{m^2}) \frac{ B_{\frac{5}{2},s-3}(R^2 |Z(\Gamma)|^2  , R^2 \sqrt{\Delta(\Gamma)})}{\Delta(\Gamma)^{\frac{s-3}{2}}}   e^{2\I\pi \langle \Gamma ,a\rangle }  \ .  \eea
Setting $d=7$, $s=\tfrac52$, \eqref{expE8s3}  reproduces \eqref{r4largestrong}, taking into account the normalization
factor from \eqref{eisr4}. Similarly, for $d=7$, $s=\tfrac92$, \eqref{expE8s3} reproduces
\eqref{d4r4largestrong}, taking into account the normalization
factor from \eqref{eisd4r4}. It is worth noting that the branches $abc)$, despite carrying
explicit dependence on the NUT scalar $b$, do contribute to the Abelian Fourier
coefficients as well for generic $s$. Indeed, we know that  the completed Eisenstein series
\be
\cE^{\star E_8}_{s\Lambda_8}=\xi(2s)\,\xi(2s-5)\,\xi(2s-9)\,\xi(4s-28)\, \cE^{E_8}_{s\Lambda_8}\ ,
\ee
is invariant under $s\mapsto \tfrac{29}{2}-s$, as required by the Langlands functional identity,
so the contributions of $abc)$ must include the images of the terms present in \eqref{expE8s3} under this symmetry, possibly along with other terms which are by themselves invariant under $s\mapsto \tfrac{29}{2}-s$.\footnote{It can be seen that additional contributions to the Abelian Fourier coefficients satisfying to $I_4^\prime(\Gamma)=0$ do in fact occur for generic values of $s$, by applying the same  argument as in  Section \ref{sec_decomp} to $\cE^{\star E_8}_{s\Lambda_8}$, using its known weak coupling expansion \cite[A.70]{Pioline:2015yea}.} The representation associated to this function for generic values of $s$ implies that it must admit non-zero generic Abelian Fourier coefficients (with $I_4(\Gamma)\ne 0$) \cite{Bossard:2015foa}. The fact that the contributions of branches $def)$ already reproduce 
the results \eqref{r4largestrong} and \eqref{d4r4largestrong} shows that the 
Abelian contributions  of branches $abc)$ vanish for the values $s=\tfrac52$ and $s=\tfrac92$
relevant for the minimal and next-to-minimal theta series.

To explain how this can be the case, let us now compare the contributions of the branches $def)$ with the
known constant terms from Langlands'  formula (see e.g. \cite[A.69]{Pioline:2015yea}):
\bea  
\int \de^{56} \hspace{-1mm} a \, \de b \; \cE^{E_8}_{s\Lambda_8}  &=& R^{4s} 
+ \frac{\xi(2s-1)}{\xi(2s)} R^{2s+1}Ê\cE^{E_7}_{(s-\scalebox{0.6}{$\frac12$})\Lambda_7} + \frac{ \xi(2s-6)\xi(2s-11)}{\xi(2s)\xi(2s-5)} R^{12}Ê\cE^{E_7}_{(s-3)\Lambda_1}\nn \\
&& \frac{ \xi(2s-18) \xi(2s-14) \xi(2s-10) \xi(4s-29)}{\xi(2s) \xi(2s-5) \xi(2s-9)  \xi(4s-28)}  R^{30-2s} \cE^{E_7}_{(s-5)\Lambda_7} \nn \\
&&  + Ê\frac{\xi(2s-19) \xi(2s-23) \xi(2s-28) \xi(4s-29) }{\xi(2s) \xi(2s-5) \xi(2s-9) \xi(4s-28)}R^{58-4s} \ . \eea
The first three terms agree with the constant terms arising from the branches f), e) and d), respectively. In all but the first term, the denominator contains a factor of  $\xi(2s)$ which ensures that all contributions vanish at $s=0$, in agreement with the fact that $\cE^{E_8}_{0\Lambda_8}=1$. Similarly, the 
factor of $\xi(2s-5)$ in the denominator of all but the first two terms ensures that only the first
two contributions remain  when $s=\frac{5}{2}$ (the value relevant for the minimal theta series). It is worth noting that this factor appears in the full contribution of 
branch e), not only its constant term. The fact that a factor of $\xi(2s-9)$ appears in all 
constant terms but for the first three suggests that the branches abc) would similarly 
carry a similar factor in the denominator, at least for the part
which contributes to  Abelian Fourier coefficients, and would explain why they appear not
contribute to the minimal and next-to-minimal theta series. 
It would be nice to
demonstrate that our formula reproduces completely the Abelian Fourier coefficients for $s=\tfrac{9}{2}$ by an explicit computation of the branches abc), though this is beyond the scope of the present work.


\subsection{Analogous computation for $E_7$}

We now apply a similar analysis to the maximal parabolic Eisenstein series $\cE^{E_{7}}_{s \Lambda_{7}}$, represented as a constrained Epstein series
\be 
\label{LEpsteinE7} 
\cE^{E_{7}}_{s \Lambda_{7}}  =  \frac{1}{2\zeta(2s)} \sum_{\substack{ \Gamma \in \IZ^{56}_* \\ 
\Gamma\times \Gamma= 0}} |Z(\Gamma)|^{-2s}  \ .
\ee
Here, the identity \eqref{LEpsteinE7} is a consequence of the known fact that the set of 
primitive rank one vectors in  $\IZ^{56}$ has a single orbit under 
$E_7(\IZ)$ \cite{krutelevich2007jordan}. 
The Fourier expansion of \eqref{LEpsteinE7} 
can be computed similarly as in the previous section, using the same decomposition of $\Gamma = ( q_0 , q,p,p^0)$ with 
\bea |Z(\Gamma)|^2 &=& R^{-3 } \scal{Êq_0 + \tr a q  + \tr a\times a p - \det a p^0 }^2 + R^{-1}Ê|v(q+ 2 a \times p - a\times a p^0)|^2 \nn \\
&& \hspace{40mm}Ê + R|\vI( p- a p^0)|^2 + R^3 (p^0)^2 \ , \eea  
The constraint $\Gamma \times \Gamma = 0$ decomposes as
\be  q_0 p  = q\times q \ , \quad 4 q \times ( p\times y) = p \tr q y - q_0 p^0 y \  , \quad 3 q_0 p^0 = -  \tr q p  \ , \quad p \times p = - p^0 q \ , \ee
where $y$ in an arbitrary element of the Jordan algebra. As in the previous section we shall solve this equation in various branches. The generic branch  $a): p^0\ne 0$ gives 
\be q = - \frac{p\times p}{p^0} \ ,\qquad  q_0 = \frac{\det p}{(p^0)^2} \ . \ee
for any $p^0$ and $p$ such that $q$ and $q_0$ are integers.  
The invariant bilinear form then reduces to 
\be |Z(\Gamma)|^2 = R^{-3 } \Scal{Ê \frac{\det(p-ap^0)}{(p^0)^2}}^2   + R^{-1}Ê|v\scal{Ê\tfrac{(p-ap^0)\times (q-ap^0)}{p^0}}|^2 + R|\vI( p- a p^0)|^2 + R^3 (p^0)^2 \ , \ee 

The second branch b) corresponds to $p^0= 0, p\ne 0$. Because $p\times p=0$ one can always find an element of $E_6(\IZ)$ to bring $p$ to a canonical diagonal form in which only the first entry is non-zero \cite{krutelevich2007jordan}. The complete solution is then
\be  
p = \left( \begin{matrix} p^1 & 0 & 0\\0&0&0\\0&0&0\end{matrix}\right)\ , \quad q = \left( \begin{matrix} 0 & 0 & 0\\0&q_2&\chi_1\\0&\chi^*_1&q_3\end{matrix}\right) \ , \quad q_0 p^1 = q_2 q_3 - \chi_1\chi_1^* \ , \ee
where $p^1,q_2,q_3,q_0$ are integers and $\chi_1$ is an integral split octonion. Up to a Poincar\'e sum over $P_6 (\IZ) \backslash E_6(\IZ)$ the sum over these charges therefore reduce to a theta lift as in the last section, where the $SO(5,5)$ even self-dual norm is defined as 
\be ( q,q) = 2 ( q_2 q_3 - \chi_1\chi_1^*) \ . \ee
The branch c) corresponds to $p^0= p= 0, q\neq 0$ with  $q\times q=0$, and the last branch d) to
$p^0= p= q=0, q_0\neq 0$. The same method as in the previous section can be applied
to evaluate the contributions of branches $bcd)$, leading to 
\bea 
\label{expE7}
\xi(2s) \cE^{E_{7}}_{s \Lambda_{7}} &=&  \xi(2s) R^{3s}+\xi(2s-1)R^{s+1} {\cal E}^{E_6}_{(s-\frac{1}{2})\Lambda_6} + \frac{\xi(2s-5) \xi(2s-9)}{\xi(2s-4) }  R^{10-s}Ê{\cal E}^{E_6}_{(s-\frac{5}{2})\Lambda_1}  \\
&& + 2  R^{2s+\frac{1}{2}}Ê\sum_{\substack{ q\in \IZ^{27}_*\\ q\times q=0}}  \sigma_{2s-1}(q) \frac{K_{s-\frac{1}{2}}(2\pi R |v(q)|)}{|v(q)|^{s-\frac{1}{2}}} e^{2\I\pi {\rm tr} q  a } \nn \\
&& + 2 \frac{\xi(2s-5)}{\xi(2s-4) } R^{\frac{11}{2}}Ê
\sum_{\substack{ q\in \IZ^{27}_*\\ q\times q=0}} \frac{ \sigma_{2s-9}(q)}{{\rm gcd}q^{s-\frac{13}{2}}} \cE^{D_5}_{(s-\frac{5}{2})\Lambda_1}( \upsilon_q) \frac{K_{s-\frac{9}{2}}(2\pi R |v(q)|)}{|v(q)|^{2}} e^{2\I\pi {\rm tr} q  a } \nn \\
&&\hspace{-15mm}
+2 \frac{R^{s+5}}{\xi(2s-4) }  \hspace{-4mm}  \sum_{\substack{ q\in \IZ^{27}\\ q\times q\ne 0, \, \det \! q=0 }} \hspace{-2mm} 
\sum_{m|q} m^{2s-1} \sigma_{2s-5}( \tfrac{q\times q}{m^2}) \frac{ B_{2,s-\frac{5}{2}}(R^2 |v(q)|^2,R^2|\vI(q\times q)|)} {|\vI(q\times q)|^{s-\frac{5}{2}} } e^{2\I\pi {\rm tr} q  a } \nn \\
&& \hspace{-15mm} 
+ \sum_{n>0} \int_0^\infty \frac{\de t}{t^{1+s}}   \hspace{-6mm}   \sum_{\substack{ p\in \IZ^{27}\\ \frac{p\times p}{n}\in \IZ^{27}, \frac{\det p}{n^2}\in \IZ}}  \hspace{-6mm}  e^{-\frac{\pi}{t} \scal{R^{-3} \scal{Ê \frac{\det(p+an)}{n^2}}^2   + R^{-1}Ê|v(Ê\frac{(p+an)\times (p+an)}{n})|^2 + R|\vI( p+an)|^2 + R^3 n^2}} \nn \,Ê,  
\eea
where the last line corresponds to the generic branch a), which we do not know how
to compute.
For $d=6$, $s=2$, the first four lines in \eqref{expE7} reproduce \eqref{r4largestrong}, taking into account the normalization
factor from \eqref{eisr4}. Similarly, for $d=6$, $s=4$, the first four lines in  \eqref{expE7} reproduce
\eqref{d4r4largestrong}, taking into account the normalization
factor from \eqref{eisd4r4}. The comparison with Langlands constant term 
formula  \cite[A.66]{Pioline:2015yea}
\bea \xi(2s) \,\cE^{E_{7}}_{s \Lambda_{7}} &=&  \xi(2s) \,R^{3s}+\xi(2s-1)\,R^{s+1} \,{\cal E}^{E_6}_{(s-\frac{1}{2})\Lambda_6} + \frac{\xi(2s-5) \,\xi(2s-9)}{\xi(2s-4) } \, R^{10-s}Ê\,{\cal E}^{E_6}_{(s-\frac{5}{2})\Lambda_1}  \nn \\
&& +\frac{ \xi(2s-9)\, \xi(2s-13) \, \xi(2s-17)}{\xi(2s-4)\, \xi(2s-8)} R^{27-3s} + \mathcal{O}(e^{-2\pi R} ) \ , \eea 
suggests that the contribution of branch a) should carry a factor of  $\xi(2s-4)\xi(2s-8)$ in the denominator, which would explain why it does not contribute  for the values $s=2,4$ associated to 
the minimal and the next-to-minimal representation of $E_7$. In fact, an analysis of the 
weak coupling limit of $\cE^{E_{7}}_{s \Lambda_{7}}$ for generic values of $s$ using 
\cite[A.67]{Pioline:2015yea} suggests that the contribution of branch a) to the Fourier coefficients 
supported on non-generic charges with $\det q=0$ is determined from
\eqref{expE7} by demanding the invariance of the completed Eisenstein series $\cE^{\star,E_{7}}_{s \Lambda_{7}}=
\xi(2s)\, \xi(2s-4)\, \xi(2s-8)\, \cE^{E_{7}}_{s \Lambda_{7}}$ under $s\mapsto 9-s$, while 
additional  Fourier coefficients supported on generic charges with $\det q\ne 0$ appear for generic 
values of $s$.

\subsection{Analogous computation for $E_{d+1}$}
More generally, the complete Fourier expansion of the  constrained lattice sum for $E_{d+1}(\IZ)$ can be computed similarly as in the previous sections for $2\le d\le 5$, by
splitting $\mathcal{Q}\in  M^{E_{d+1}}_{\Lambda_{d+1}}$ according to the graded decomposition relevant for the decompactification limit \cite[(4.26)]{Bossard:2015foa},  
\begin{align}
\label{1ChargeGrad}  
 \mathfrak{e}_{d+1(d+1)} &\cong {\bf R}_{\Lambda_{d}}^{\ord{d-8}} \oplus \scal{ \mathfrak{gl}_1 \oplus   \mathfrak{e}_{d(d)}}^\ord{0} \oplus  \overline{\bf R}_{\Lambda_{d}}^{\ord{8-d}} \ , \nn\\
{\bf R}_{\Lambda_{d+1}} &\cong  {\bf R}_{\Lambda_1}^\ord{d-7} \oplus {\bf R}_{\Lambda_{d}}^\ord{1} \oplus {\bf 1}^\ord{9-d} \ ,\nn\\
{\bf R}_{\Lambda_1} &\cong  {\delta}_{d,5}^\ord{-4}   \oplus  {\bf R}_{\Lambda_2}^\ord{d-6} \oplus {\bf R}_{\Lambda_1}^\ord{2} \ ,  
\end{align}
where $ {\delta}_{d,5}^\ord{-4} $ denotes a singlet which appears only for $d=5$. 
Decomposing $\mathcal{Q} = ( Q,\Gamma,m) $ with $Q\in  (M^{E_{d}}_{\Lambda_1})^\ord{d-7}, \Gamma\in  (M^{E_{d}}_{\Lambda_d})^\ord{1}$ and $m\in{\bf 1}^\ord{9-d}$, 
The constrained Epstein series \eqref{LEpstein} can then be written as 
\bea && 
2  \xi(2s)\cE^{E_{d+1}}_{s \Lambda_{d+1}}   \\
&=& \hspace{-4mm}\sum'_{\substack{ Q\in M^{E_d}_{\Lambda_1}, \Gamma \in M^{E_d}_{\Lambda_d} , m\in\IZ, \\
Q^2=Q^2|_{{R}_{2\Lambda_1}}, Q \Gamma |_{{R}_{\Lambda_2}} = 0,\\
 \Gamma \times \Gamma = m Q}}  \int_0^\infty  \frac{ \de t}{t^{1+s}} e^{-\frac{\pi}{t} \scal{R^{2\frac{d-9}{8-d}} (Ê m + \langle a , \Gamma \rangle + \frac{1}{2} \langle a , Q a \rangle )^2 + R^{\frac{2}{d-8}} |Z(\Gamma + Q a)|^2 + R^{2\frac{7-d}{8-d}} |V(Q)|^2 }}\nn \ .  \eea
Here we use Bourbaki label's for the weights $\Lambda_i$ of $E_d$ (even when $E_d$ is a classical
group), and $\Lambda_d$ for $E_3$ is $\Lambda_2+\Lambda_3$ (we refer to \cite{Bossard:2015foa} where this notation was introduced for more details). The sum over $(0,\Gamma,m)$ can be  carried out by Poisson resummation over $m$ and the sum over elements including a non-zero $Q$ can be computed similarly as in the previous sections. A non-zero $Q$ (which is a null vector of $SO(5,5)$ for $d=5$ and a generic element of $M^{E_d}_{\Lambda_1}$ otherwise) is in the $E_d(\IZ)$ orbit of a canonical element of degree $4$ ($Q=(0,0,n)$) in the decomposition 
\bea \mathfrak{e}_{d(d)} &\cong& {\bf S}_+^\ord{d-9} \oplus \scal{ \mathfrak{gl}_1 \oplus   \mathfrak{so}(d-1,d-1)}^\ord{0} \oplus \overline{\bf S}_+^\ord{9-d} \ , \nn\\
{\bf R}_{\Lambda_d} &\cong& {\bf S}_-^\ord{d-7} \oplus {\bf V}^\ord{2}  \ , \nn\\
{\bf R}_{\Lambda_1} &\cong& {\delta}_{d, 5}^\ord{-4}  \oplus {\bf S}_+^\ord{d-5} \oplus {\bf 1}^\ord{4} \ . \eea
The representation ${\bf R}_{\Lambda_2}^\ord{d-6}$ includes ${\bf S}_-^\ord{d-3}$ as its highest grade component,  so that the constraint $Q \Gamma |_{{\bf R}_{\Lambda_2}} = 0$ implies that $\Gamma = ( 0,q)$ with respect to this decomposition. The constraint in the component ${\bf R}_{\Lambda_1}^\ord{2}$ eventually implies  that $2m n = (q,q)$. 

Using the definitions 
 \be 
 | V(Q)|^2 = e^{2\phi} n^2 \ , \qquad |Z(\Gamma)|^2 = e^{\phi} g(q,q) \ , \qquad  t  = \frac{e^\phi}{R^{\frac{2}{8-d}}  \rho_2} \, , 
 \ee
the sum can be recognized as the Poincar\'e sum over $P_1\backslash E_d(\IZ)$ of the integral over the strip $\cH/\IZ$ of the Siegel--Narain theta series of signature $SO(d,d)$, multiplied by  $\mathcal{E}(s-\tfrac{d-2}{2},\rho)-\rho_2^{\; s-\frac{d-2}{2}}$. The Fourier coefficient for which $q$ is a null vector corresponds to the sum over primitive $Q$ and $\Gamma$ satisfying to $\Gamma \times \Gamma = 0$ and $Q \Gamma |_{{\bf R}_{\Lambda_2}} = 0$.  Therefore the Poincar\'e sum over $P_1\backslash E_d(\IZ)$ and null vectors $q$ can be rewritten as a Poincar\'e sum over $P_d\backslash E_d$ and the primitive rank 1 charges $Q\in {\bf R}_{\Lambda_1}^\ord{2}$ in the corresponding decomposition under
$E_{d-1}$. Using the change of variable
\be 
e^{2\phi}  = e^{2\upsilon} \Bigl(\frac{|Z(\Gamma)|}{{\rm gcd} \Gamma} \Bigr)^\frac{4}{10-d}\ , 
\ee 
where $ e^{2\upsilon}$ is the character of $P_{1} \subset E_{d-1}$, one obtains after some work 
\bea
\label{expEd}
 \xi(2s)\, \cE^{E_{d+1}}_{s \Lambda_{d+1}} &=& \xi(2s)R^{2\frac{9-d}{8-d} s} +\xi(2s-1)R^{\frac{2}{8-d}s+1} \mathcal{E}^{E_d}_{(s-\frac{1}{2}) \Lambda_d}  \\ 
 &&+\frac{\xi(2s-d+1) \xi(2s-2d+3)}{\xi(2s-d+2) } R^{-2\frac{7-d}{8-d}s+ 2(d-1)} \mathcal{E}^{E_d}_{(s-\frac{d-1}{2})\Lambda_1}
\nn \\ 
&& \qquad + \,2 R^{\frac{10-d}{8-d}s +\frac{1}{2}}  \sum_{\substack{\Gamma \in M^{E_d}_{\Lambda_d*}\\ \Gamma \times \Gamma = 0}} \sigma_{2s-1}(\Gamma)\frac{ K_{s-\frac{1}{2}}( 2\pi R|Z(\Gamma)|)}{|Z(\Gamma)|^{s-\frac{1}{2}}} e^{2\pi \I \langle \Gamma, a\rangle } \nn \\
&&\hspace{-25mm}+ 2\frac{\xi(2s-d+1)}{\xi(2s-d+2) } R^{-\frac{6-d}{8-d}s+d-\frac{1}{2}}\hspace{-2mm}\sum_{\substack{\Gamma \in M^{E_d}_{\Lambda_d*}\\ \Gamma \times \Gamma = 0}} \hspace{-1mm} \frac{ \sigma_{2d-3-2s}(\Gamma) }{ {\scalebox{0.8}{$(\mbox{gcd } \Gamma)^{\frac{2(2s-d+1)}{d-10}}  
$}}}    \cE^{E_{d-1}}_{(\scalebox{0.7}{$s$-$\frac{d-1}{2}$}) \Lambda_1}\hspace{-1mm}(g_\Gamma) \frac{K_{s-d+\frac{3}{2}}(2\pi R | Z(\Gamma)|)}{ {\scalebox{0.9}{$|Z(\Gamma)|^{\frac{\scalebox{0.6}{$(d-6)(s-d+{\scalebox{0.9}{$\frac{3}{2}$}})+2(d-2) $}}{10-d}} $}}} e^{2\I\pi\langle \Gamma , a\rangle }\nn \\
 &&\hspace{-25mm}+\frac{ 2  R^{\frac{2s}{8-d}+d-1}}{\xi(2s-d+2) } \sum_{\substack{\Gamma \in M^{E_d}_{\Lambda_d*} \\ \Gamma \times \Gamma \ne  0}} \sum_{m|\Gamma} m^{2s-1} \sigma_{2s-d+1}(\tfrac{\Gamma\times \Gamma}{m^2})  \frac{B_{\frac{d-2}{2},s-\frac{d-1}{2}}\scal{R^2 | Z(\Gamma)|^2,R^2 \sqrt{\Delta(\Gamma)}}}{\Delta(\Gamma)^{\frac{2s-d+1}{4}}} e^{2\I\pi\langle \Gamma , a\rangle } \ .   \nn \eea
For $d=6$ and $d=7$, \eqref{expEd} reduces to the  results \eqref{expE7} and \eqref{expE8s3}, up to the contributions of branches abc) and a), respectively, while for $d\leq 5$, \eqref{expEd} provides the complete Fourier expansion of $\cE^{E_{d+1}}_{s \Lambda_{d+1}}$. For $s=\tfrac{d-2}{2}$ and
$s=\tfrac{d+2}{2}$, taking into account the normalization
factors from \eqref{eisr4}, \eqref{expEd} reproduces  \eqref{r4largestrong} and  \eqref{d4r4largestrong}, respectively. Unlike the cases $d=6$ and $d=7$ discussed
previously, the Fourier expansion \eqref{expEd} is complete for arbitrary values of $s$ away from
the poles.


\section{Discussion \label{sec_discuss}}

In this note, we have analyzed the asymptotics of the exact $\nabla^4\cR^4$ coupling $\cE_{\scalebox{0.6}{$(1,0)$}}^{\scalebox{0.6}{$(D)$}}$ in type II string theory on $\IR^{1,D-1}\times T^{10-D}$, in the limit where the radius $R$ of one circle inside  the internal torus becomes very large. Beyond the known power-like terms, this expansion exposes an infinite series of exponentially suppressed contributions of order $e^{-2\pi R \cM(\Gamma)+2\pi\I\langle \Gamma,a\rangle}$, where $\Gamma$ runs over  the charge lattice, $\cM(\Gamma)$ is the mass of a BPS state of charge $\Gamma$ and $a$ are the holonomies of the $D+1$-dimensional Maxwell fields around the circle,
along with $\cO(e^{-2\pi R^2})$ corrections from Taub-NUT instantons when $D=3$. Using the property that only 1/4-BPS and 1/2-BPS charge vectors $\Gamma$ contribute to the sum \cite{Green:2010wi,Green:2011vz} and the  tensorial differential equations implied by supersymmetry  \cite{Bossard:2014lra,Bossard:2014aea}, we have shown that the complete large radius expansion of $\cE_{\scalebox{0.6}{$(1,0)$}}^{\scalebox{0.6}{$(D)$}}$ (up to non-Abelian Fourier coefficients when $D=3$) is determined unambiguously from its perturbative expansion in the string coupling constant. Remarkably, we find 
that the contribution of 1/4-BPS charge vectors is proportional
to the helicity supertrace $\Omega_{12}(\Gamma)$, which counts (with signs) 1/4-BPS states in dimension $D+1$. This result supports the general philosophy that BPS-saturated couplings in
dimension $D$ can provide a useful bookkeeping device for precision counting of BPS 
black hole microstates in dimension $D+1$. In the present case however, the power-like growth of 
the BPS index $\Omega_{12}(\Gamma)$ at large $|\Gamma|$ prevents any direct comparison with 
the entropy of macroscopic black holes. 

Clearly, it would be important to investigate the next coupling
in the low-energy expansion, namely the $\nabla^6\cR^4$ coupling in $D=3$, which is expected to count 1/8-BPS black holes in dimension $4$. A tantalizing hint that 1/8-BPS instanton contributions will turn out to be proportional to the BPS index $\Omega_{14}$ is the observation that the same 
weight $-2$, index $1$ Jacobi form which determines $\Omega_{14}$ \cite{Maldacena:1999bp}
also determines the Fourier expansion of the Kawazumi--Zhang invariant \cite{Pioline:2015qha}, 
which enters in the $\nabla^6 \cR^4$ coupling \cite{D'Hoker:2013eea}. 
Moreover, the differential equation satisfied by the $\nabla^6\cR^4$ coupling in $D=3$ is associated
to the same nilpotent orbit \cite{Bossard:2015uga,Bossard:2015foa} which describes 1/8-BPS
black holes in $D=4$ \cite{Bossard:2009at}. In more physical terms this means that the only instantons that contribute to this coupling according to the supersymmetry constraint are precisely the ones that can be interpreted as 1/8 BPS black holes in four dimensions, or more generally 1/8 BPS Taub-NUT black holes.  It would be very interesting to
see if this connection can be explicited within existing proposals for the exact $\nabla^6 \cR^4$ coupling \cite{Green:2005ba,Basu:2007ck,Green:2014yxa,Pioline:2015yea,Bossard:2015foa}.

As for the 1/2 and 1/4-BPS couplings studied in this work, it would be highly desirable to compute
the contributions of 1/2- and 1/4-BPS black holes from first principles, and reproduce the detailed
dependence on the radius $R$ and central charges $Z(\Gamma), \Delta(\Gamma)$ found in  
\eqref{d4r4largestrong} and \eqref{r4largestrong}, including the `multi-covering' effects which 
appear when $\Gamma$ is not a primitive vector. In particular, it would be very interesting to
interpret the Eisenstein series $\cE^{\scalebox{0.7}{$E_{d-1}$}}_{\scalebox{0.65}{$\frac32\Lambda_1$}}(g_\Gamma)$ under
the Levi subgroup of the stabilizer of $\Gamma$, possibly in terms of higher-derivative corrections
around the instanton saddle point. It would also be interesting to extract the non-Abelian 
Fourier coefficients from gravitational Taub-NUT instantons. Finally, from the mathematical 
point of view it would be very interesting to see  the Fourier expansion \eqref{d4r4largestrong}
emerge from a general adelic approach, and understand how it generalizes to Fourier
expansion with respect to other choices of maximal parabolic subgroups. 

\bigskip

{\noindent \it Acknowledgements:} We are grateful to C. Cosnier-Horeau, D. Persson, and in particular to A. Kleinschmidt, for valuable discussions, to the organizers of the program "Automorphic forms, mock modular forms and string theory" at the Simons Center for Geometry and Physics for providing a stimulating
atmosphere which made this work possible, and to the Simons Center for its warm hospitality.

\medskip

\appendix

\section{Decompactification limits of certain modular integrals}

In this section we compute the decompactification limits of the modular integrals
appearing in \eqref{dfourrfournew} and \eqref{dfourrfour2}, using the unfolding
method. Since these  modular integrals can be expressed in terms of maximal parabolic 
Eisenstein series of $O(d,d)$, this provides the complete Fourier expansion of these
Eisenstein series with respect to the maximal parabolic subgroup $P_{1}$.

\subsection{Genus one modular integral}

The genus one regularized modular integral
\be
\label{defIs}
\cI_{d,1}(s) = \RN\, \int_{\cF_1} \de \mu_1\, \cE^\star(s,\rho)\, \Gamma_{d,d,1}(\rho)\ ,
\ee
where $\cE^\star(s,\rho)=\xi(2s) \cE(s,\rho)$ is the completed non-holomorphic Eisenstein
series of $SL(2,\IZ)$ and $\RN$ denotes the regularization prescription of \cite{MR656029}. It  
is  proportional to the parabolic Eisenstein series attached
to the vector representation of $O(d,d)$, \cite{Green:2010wi,Angelantonj:2011br}:
\be
\cI_{d,1}(s) = 2\xi(2s)\, \xi(2s+d-2)\, 
\cE^{D_d}_{(s+\frac{d-2}{2})\Lambda_1}\ .
\ee 
To compute its  limit when the radius $R_s$ of one circle inside $T^d$ becomes large in string units, 
we perform a Poisson resummation on the corresponding momentum, and write 
\be
\Gamma_{d,d,1}(\rho) = R_s\, \rho_2^{\frac{d-1}{2}}\,  \sum_{m,n\in\IZ} e^{-\frac{\pi R_s^2 |m-n\rho|^2}{\rho_2}}
\hspace{-2mm}\sum_{Q\in\IZ^{d-1,d-1}} \hspace{-2mm} q^{\frac12 p_L^2(Q-n a)} \bar q^{\frac12 p_R^2(Q-n a)
+2\pi\I m (Q,a)-\I\pi m n (a,a)}\ ,
\ee 
where $a\in\IR^{d-1,d-1}$ parametrize the nilpotent radical in the parabolic subgroup $P_{1}$
with Levi $\IR^+\times O(d-1,d-1)$, while $R_s$ parametrizes the center of $P_{1}$. The
doublet  $(m,n)$ transforms linearly under  $SL(2,\IZ)$. The contribution from $(m,n)=0$
produces $R_s \cI_{d-1}(s)$, while the sum over $(m,n)\neq (0,0)$ can be restricted to
the orbit representatives $(m,0)$ with $m\neq 0$, at the cost of extending the integration
domain\footnote{Here we work formally, ignoring the regularization which is necessary to make sense
of the integral \eqref{defIs}. This issue only affects the constant terms, which were dealt more 
rigorously in \cite[\S 3.3]{Angelantonj:2011br}.}
from $\cF_1$ to the strip $\cH/\IZ$:
\be
\cI_{d,1}(s) = 
R_s \cI_{d-1}(s) + R_s \sum_{m\neq 0} \int_{\cH/\IZ} \de\mu_1\, \cE^\star(s,\rho) 
\hspace{-2mm}\sum_{m\neq 0 \atop Q\in\IZ^{d-1,d-1} } \hspace{-2mm}e^{-\frac{\pi m^2 R_s^2}{\rho_2}+2\pi\I m (Q,a)}\, 
 q^{\frac12 p_L^2(Q)} \bar q^{\frac12 p_R^2(Q)}\ .
\ee 
We now substitute the Chowla--Selberg formula for 
$\cE(s,\rho)$,
\be
\label{chowla}
\cE(s,\rho) = \xi(2s)\, \rho_2^s + \xi(2s-1)\, \rho_2^{1-s} + 
2\,\sum_{N\neq 0}\, |N|^{s-\frac12}\,\sigma_{1-2s}(|N|)\,\sqrt{\rho_2}
K_{s-\frac12}(2\pi|N|\rho_2)\,e^{2\pi\I N \rho_1}\ .
\ee
For $Q=0$, only the first terms contribute to the integral over $\rho_1$, leading
to constant terms proportional to $R_s^{2s+2-2}$ and $R_s^{d-2}$. For $Q\neq 0$,
the integral over $\rho_1$ picks up the term with $2N=p_L^2(Q)-p_R^2(Q)$, and 
the integral over $\rho_2$ is of Bessel type. After rescaling $Q\to Q/m$, we obtain
\be
\label{Idsdec}
\begin{split}
\cI_{d,1}(s)=&  R_s\, \cI_{d-1}(s) + 2\xi(2s)\,\xi(2s+d-2)\, R_s^{2s+d-2}
 + 2\xi(2s-1)\,\xi(2s-d+1)\, R_s^{d-2s}
 \\
  +& 4\, \xi(2s)\, R_s^{s+\frac{d-1}{2}}  \sum_{Q\in\IZ^{d-1,d-1}_*\atop p_L^2-p_R^2=0} 
 \frac{\sigma_{2s+d-3}(Q)}{(p_L^2+p_R^2)^{\frac{2s+d-3}{4}}}\,
 K_{s+\frac{d-3}{2}}\left(2\pi R_s \sqrt{p_L^2+p_R^2}\right)\, e^{2\pi\I (Q,a)}
\\
 + & 4\, \, \xi(2s-1)\, R_s^{\frac{d+1}{2}-s}
\hspace{-2mm} \sum_{Q\in\IZ^{d-1,d-1}_*\atop p_L^2-p_R^2=0} 
\frac{\sigma_{d-1-2s}(Q)}{(p_L^2+p_R^2)^{\frac{d-1-2s}{4}}},
K_{s-\frac{d-1}{2}}\left(2\pi R_s \sqrt{p_L^2+p_R^2}\right)\, e^{2\pi\I (Q,a)}
\\
 &\hspace*{-1cm} 
+ 4 R_s^{d-1}\!\!\!\!\!\ \hspace{-1mm} \sum_{Q\in\IZ^{d-1,d-1}_*\atop p_L^2-p_R^2\equiv 2 N\neq 0}
\, \sum_{m\in \IN_* \atop 
m|Q} \frac{|N|^{s-\frac12}}{m^{2s+1-d}}\,\sigma_{1-2s}\left(\frac{|N|}{m^2}\right)
B_{\frac{d-2}{2},s-\frac12}\left(R_s^2 (p_L^2+p_R^2), |N| R_s^2\right)\, e^{2\pi\I (Q,a)}
\end{split}
\ee
where $p_L\equiv p_L(Q), p_R\equiv p_R(Q)$, and $B_{s,\nu}(x,y) $ is the integral defined in \eqref{defBsnu}. Setting $s=2$ produces \eqref{d4r41loopR}.

\subsection{Genus $h$ modular integral}

We now consider the modular integral 
\be
\label{defIh}
\cI_{d,h}  =  \RN \int_{\mathcal{F}_h} \, \de\mu_h\, \Gamma_{d,d,h}
\ee
over a fundamental domain $\cF_h$ of the degree $h$ Siegel upper half-plane $\cH_h$.
The case of interest in this work is $h=2$, but our treatment will work for any $h\geq 1$. 
As in \eqref{defIs}, the integral diverges if $d\geq h+1$, and must be regularized. A regularization
scheme was described in detail for $1\leq h\leq 3$ in \cite{Florakis:2016boz}, and is easily generalized to arbitrary $h$. The regularized integral can then be expressed as a residue at $s=\tfrac{d-3}{2}$ of the maximal parabolic Eisenstein series $\cE^{D_d}_{s\Lambda_h}$ attached to the rank $h$ antisymmetric representation \cite{Pioline:2014bra,Florakis:2016boz}, or as a residue at $s=h$ of  the maximal parabolic
Eisenstein series attached to the spinor representations \cite{Obers:1999um}. 

Regardless of its connection to Eisenstein series, its limit as the radius $R_s$ of one circle in $T^d$ goes to infinity can be computed as in the previous subsection using
the orbit method. After Poisson resummation over the momenta, one finds
\begin{eqnarray} 
\cI_{d,h} 
 &=&   \int_{\mathcal{F}_h}  \de\mu_h\, |\Omega_2|^{\frac{d-1}{2}}\, R_s^hÊ\hspace{-5mm}  \sum_{\substack{Q^i\in (\IZ^{d-1,d-1})^{\otimes h}\\ (m_i,n^i)\in \IZ^{2h}}} 
 \hspace{-4mm} 
e^{\I\pi \Omega_{ij}\, p_L(Q^i - a n^i) p_L(Q^j - a n^j ) 
 -  \I\pi \bar\Omega_{ij}\, p_R(Q^i - a n^i) p_R(Q^j - a n^j ) }
 \nn\\
 &&   \hspace{25mm} \times e^{ -\pi R_s^2 \Omega_2^{-1ij} ( m_i + \Omega_{ik} n^k) ( m_j + \bar \Omega_{jl} n^l) +2\I\pi ( m_i Q^i , a) -\I\pi m_i n^i  (  a , a) }\nn  
 \ .
 \end{eqnarray} 
The zero orbit $(m_i,n^i)=0$ leads to $R_s^h\, \cI_{d-1,h}$, while the other orbits can be reduced
to the representatives  $(m_i,n^i)=( \delta_i^h m , 0)$ with $m\in \IN^*$, at the expense of extending
$\cF_h$ to a fundamental domain of the stabilizer $Sp(2h-2,\IZ)\ltimes (\IZ^{2(h-1)+1})
\subset Sp(2h,\IZ)$ \cite[\S 4.5.2]{Pioline:2014bra}. For notational simplicity we shall keep the same notations for the smaller symplectic group, so $i$ is ranging now from $1$ to $h-1$ and $\Omega$ in $\cI_{d,h}$  
decomposes as 
\begin{equation} 
\Omega = \left (\begin{array}{cc} \Omega_{ij} & v_j + \Omega_{jk} u^k \, \\ v_i + \Omega_{ik} u^k  &\,   \Omega_{kl} u^k u^l + i t +\sigma\end{array} \right)   \ . 
\end{equation}
where $\Omega_{ij}$ on the r.h.s. belongs to $\cH_{h-1}$. 
Since the measure on $\cH_h$ decomposes as 
\begin{equation} 
\de\mu_h =  \de\mu_{h-1}
\frac{\de t}{t^{h+1}}\, \de^{h-1} u \,\de^{h-1} v \, \de\sigma \ ,
\end{equation}
one gets 
 \begin{eqnarray} 
\cI_{d,h} &=&R_s^{h}\, \cI_{d-1,h} + R_s^{h}\, \
 \int_{\mathcal{F}_{h-1}}\!\!\!\! \de\mu_{h-1}\, |\Omega_2|^{\frac{d-1}{2}}\,
 \int_0^\infty \!\!\!\!\frac{\de t}{t^{1+h-\frac{d-1}{2}}} \int_{[0,1]^{2h-1}} \de^{h-1} u  \,
\de^{h-1} v \, \de\sigma   \nn \\
 && 
\times \sum_{Q^i,Q \in \IZ^{d-1,d-1}}\, 
 e^{\I\pi \Omega_{ij}\, p_L(Q^i +  u^i Q) p_L(Q^j + u^j Q) 
 -  \I\pi \bar\Omega_{ij}\, p_R(Q^i + u^i Q) p_R(Q^j + u^j Q)} \nn\\
 && \times \sum_{m\in\IN^*}\, e^{
 - \frac{\pi R_s^2 m^2}{t} - \pi t |Z(Q)|^2 - 2\pi \I  v_i ( Q^i , Q )   - \pi \I  \sigma ( Q , Q ) 
 +2\pi \I m (  Q , a)} \ ,
 \end{eqnarray} 
 where $Q\equiv Q^h$. 
For $Q=0$, the integral over $u^i,v_i,\sigma$ is trivial and one recovers $\cI_{d-1,h-1}$. 
For $Q\ne 0$ the integrals over $v^i$ and $\sigma$ imply $( Q , Q ) = 0$ and $( Q , Q^i ) = 0$.
The vector $Q^i$ can then be decomposed into $Q^i = Q^i_{\bot} + r^i \frac{Q}{{\rm gcd}Q}$, with $r^i \in \IZ^{h-1}$ and $Q^i_{\bot}\in (\IZ^{d-2,d-2})^{\otimes (h-1)}$ runs over 
vectors orthogonal to $Q$,
modulo translations by $Q$.  The 
integral over $u^i$ can be unfolded to $\IR^{h-1}$ at the expense of restricting the sum over $r^i$ to 
$\IZ^{h-1}$ modulo ${\rm gcd}Q$. The dependence of the integrand on $r^i$ drops out after the Gaussian integration of the $u^i$ over $\IR$,  so the sum over $r^i$ simply produces a factor of $({\rm gcd}Q)^{h-1}$. The remaining integral over $\Omega\in\cF_{h-1}$ gives a modular integral of the form $\cI_{d-2,h-1}(v_Q)$, for the lattice of signature $(d-2,d-2)$ orthogonal to $Q$ (more precisely, the quotient of the orthogonal complement of $Q$ in $\IZ^{d-1,d-1}$ modulo $Q$). Altogether, one finds
 \begin{eqnarray} 
 \label{Idhdec}
 \cI_{d,h} &=&R_s^{h}\, \cI_{d-1,h}  +\xi(d-2h)  \,  R_s^{d-h-1}  \cI_{d-1,h-1} \nn\\
 &+&2R_s^h\,  \int_0^\infty \hspace{-2mm}\frac{\de t}{t^{1+h-\frac{d-1}{2}}} 
 \!\!\!\!\! \sum_{\substack{Q\in \IZ^{d-1,d-1}_*\\ ( Q,Q) = 0}}  
 \sum_{m\in \IN^*} \frac{({\rm gcd} Q)^{h-1}}{|Z(Q)|^{h-1}} \cI_{d-2,h-1} (v_Q)\, 
 e^{- \pi R_s^2Ê\frac{m^2}{t} - \pi t |Z(Q)|^2+ 2\pi \I  ( m Q , a)} \nn \\
  &=&
  R_s^h\,   \cI_{d-1,h} +\xi(d-2h)  \,  R_s^{d-h-1}  \cI_{d-1,h-1}  \\
 &+&2 R_s^{\frac{d-1}{2}}\hspace{-2mm} \sum_{\substack{Q \in \IZ^{d-1,d-1}_*\\ ( Q,Q) = 0}}  \hspace{-1mm}
 \sigma_{d-1-2h}(Q)\, 
({\rm gcd}Q)^{h-1}\, 
 \cI_{d-2,h-1}(v_Q)\, \frac{K_{\frac{2h+1-d}{2}}(2\pi R_s |Z(Q)|)}
 {|Z(Q)|^{\frac{d-3}{2}}} 
 e^{2\pi \I  ( Q , a)}\ , \nn
 \end{eqnarray} 
 where it is understood that $ \cI_{d,h} = 2$ when $h=0$. The modular 
 integral $\cI_{d-2,h-1}$ can be rewritten as a parabolic Eisenstein series for the Levi 
 subgroup of the stabilizer of $Q$ if
so desired. Setting $h=2$ produces
 the result stated in \eqref{d4r42loopR}.
 
\section{Differential equation of the 1/4-BPS coupling}
\label{DiffEqE8}
Supersymmetry Ward identities imply that the function $\cE_{\scalebox{0.6}{$(1,0)$}}^{\scalebox{0.6}{$(3)$}}$ satisfies 
the tensorial differential equation \cite{Bossard:2014lra}
\be 
\Gamma^{kl\, AB} \cD_B \cD_C \Gamma_{ijkl}{}^{CD} \cD_D\, \mathcal{E}_{(1,0)}^{(3)} = - 168\,  \Gamma_{ij}{}^{AB} \cD_B\, \mathcal{E}_{(1,0)}^{(3)} \  , \label{3Dd4R4} 
\ee
which implies, by integrability, \cite{Green:2010kv}
\be 
\Delta \mathcal{E}_{(1,0)}^{(3)} = - 180\,\mathcal{E}_{(1,0)}^{(3)} \ , \label{180Laplace}  
\ee
where $\cD_A$ is the covariant derivative in the Weyl spinor representation of $Spin(16)$ and $ \Gamma^{ij} = \tfrac{1}{2} [ \Gamma^i, \Gamma^j]$ are the associated 16 gamma matrices in the Weyl representation.  We refer to \cite{Bossard:2014lra} for further details on the conventions used in this section. Equation \eqref{3Dd4R4} implies that $\cD^3 \cE_{\scalebox{0.6}{$(1,0)$}}^{\scalebox{0.6}{$(3)$}}$ vanishes in the irreducible representation associated to the highest weight $\Lambda_2 + \Lambda_8$ of $Spin(16)$. Decomposing this equation with respect to $U(8)$, one obtains a $\overline{\bf 28}$ at $U(1)$ weight 3 and ${\bf 720}\oplus {\bf 63}$ at $U(1)$ weight 2. Considering this equation for an Abelian Fourier coefficient $F_\Gamma(R,t) e^{2\pi i \langle a,\Gamma\rangle}$ where $t$ parametrises the Levi subgroup $E_{7(7)}/SU(8)$ one obtains the 1/4-BPS constraint at weight 3
\be 
(\pi R)^3 Z( I_4^\prime(\Gamma))_{ij} \, F_\Gamma(R,t) = 0  \ , 
\ee
along with
\bea &&Ê \Biggl( ÊD_{ijpq}ÊD^{klpq}Ê( R \tfrac{\partial \, }{\partial R} - 10 ) - (\pi R)^2 Z_{ij}(\Gamma) Z^{kl}(\Gamma) ( ÊR \tfrac{\partial \, }{\partial R} - 6 ) \nn \\&&\hspace{10mm}
 + 2(\pi R)^2( Z_{ij}(\Gamma) Z_{pq}(\Gamma) D^{klpq} + Z^{kl}(\Gamma) Z^{pq}(\Gamma) D_{ijpq} ) \nn \\&&\hspace{20mm}
-8(\pi R)^2 (  Z_{ip}(\Gamma) Z_{jq}(\Gamma) D^{klpq} + Z^{kp}(\Gamma) Z^{lq}(\Gamma) D_{ijpq} ) \Biggr) F_\Gamma(R,t) \nn \\
&=& \tfrac{1}{28} \delta_{ij}^{kl}\Biggl( ÊD_{pqrs}ÊD^{pqrs}Ê( R \tfrac{\partial \, }{\partial R} - 10 ) - (\pi R)^2 Z_{pq}(\Gamma) Z^{pq}(\Gamma) ( ÊR \tfrac{\partial \, }{\partial R} - 6 ) \nn \\&&\hspace{10mm}
 + 10(\pi R)^2( Z_{pq}(\Gamma) Z_{rs}(\Gamma) D^{pqrs} + Z^{pq}(\Gamma) Z^{rs}(\Gamma) D_{pqrs} ) \Biggr)   F_\Gamma(R,t)  \  \label{3Dd4R4E7}Ê\eea
 at weight 2, where $D_{ijkl}$ is the complex self-dual covariant derivative on $E_{7(7)}/SU(8)$. This equation can be derived using the decomposition of \eqref{3Dd4R4}, but to obtain it we simply wrote the most general ansatz for a third order differential operator in the representation ${\bf 720}\oplus {\bf 63}$ and determined the coefficients such that the Fourier coefficients of $\cE_{\scalebox{0.6}{$(0,0)$}}^{\scalebox{0.6}{$(3)$}}$ and 
$\cE_{\scalebox{0.6}{$(1,0)$}}^{\scalebox{0.6}{$(3)$}}$ for vanishing and 1/2-BPS charges $\Gamma$ are indeed solutions. 

For a 1/2-BPS charge $\Gamma$, one can always assume that $Z_{ij}(\Gamma) = \frac{1}{\scalebox{0.6}{$2\sqrt{2}$}} \Omega_{ij} e^{3\phi} n $ for an integer $n$ and a symplectic form $\Omega_{ij}$ (with $\Omega_{ik}Ê\Omega^{jk} = \delta_i^j$) defining the embedding of $E_{6(6)}\subset E_{7(7)}$ such that 
\be D_{ijkl} = - \tfrac{1}{8} \Omega_{[ij} \Omega_{kl]} \partial_\phi + \tfrac{3i}{\sqrt{2}} e^{-2\phi} \Omega_{[ij} V_{kl]}{}^I \partial_I + \cD_{ijkl} \ , 
\ee
where $V_{ij}{}^I=V(y)_{ij}{}^I$ is the coset representative of $E_{6(6)}/Sp(4)$, $\cD_{ijkl} $ is the covariant derivative on the same space, and $\partial_I$ is the derivative with respect to the 27 axions $\alpha^I$. Decomposing \eqref{3Dd4R4E7} accordingly one obtains 
\bea
&& e^{-2\phi}\Scal{  \tfrac{1}{2}  V^{pq \, I}{}\partial_I \cD_{pq}{}^{kl} +V^{kl \, I}{}\partial_I \scal{ 2 + \tfrac{1}{6} \partial_\phi}  - 2 ( \pi R e^{3\phi}n)^2   V_{ij}{}^I \partial_I }  Ê( R \tfrac{\partial \, }{\partial R} - 10 )  F_\Gamma(R,t)  = 0\ , \nn\\ 
&&\tfrac{1}{4} e^{-4\phi} V^\inv{}_I{}^{ij} t^{IJK} \partial_J \partial_K \,   Ê( R \tfrac{\partial \, }{\partial R} - 10 )  F_\Gamma(R,t)  = 0\ ,  \nn \\
&&\sqrt{2} e^{-2\phi} \Scal{ V^{p[i\, I} \cD_p{}^{j]kl} -V^{p[k\, I} \cD_p{}^{l]ij} } \partial_I  Ê( R \tfrac{\partial \, }{\partial R} - 10 ) F_\Gamma(R,t)  = 0\ ,
\label{BPSCons}
\eea
and 
\begin{multline}
\Bigl( -\scal{ \tfrac{1}{18} \partial_\phi^{\; 2} + \tfrac{2}{3} \partial_\phi + \tfrac{1}{4} e^{-4\phi} V_{pq}{}^I V^{pqJ}\partial_I \partial_J   } \delta_{ij}^{kl} + \cD_{ijpq} \cD^{klpq} +\cD_{ij}{}^{kl} \scal{ 1  + \tfrac{1}{6} \partial_\phi} \hspace{20mm}\Bigr .   \\
\hspace{0mm} \Bigl . +  e^{-4\phi} \scal{ \delta_{[i}^{[k} V_{j]p}{}^I V^{l]p\, J} + V_{[i}{}^{[k\, I} V_{j]}{}^{l]\, J} } \partial_I \partial_J \Bigr) Ê( R \tfrac{\partial \, }{\partial R} - 10 ) F_\Gamma(R,t) \\
 =( \pi R e^{3\phi}n)^2  ( 2 \cD_{ij}{}^{kl} -  R \tfrac{\partial \, }{\partial R}  + 6 - \tfrac{1}{6} \partial_\phi) F_\Gamma(R,t)  \ .  \label{E6}
\end{multline}
Our conventions for $Sp(4)$ are such that all the antisymmetric indices are understood to be in symplectic traceless representations even if we do not write explicitly the projectors for brevity \cite{Bossard:2014lra}. The second equation in \eqref{BPSCons} implies that the Fourier expansion
of $ F_\Gamma(R,t)$ with respect to the axions $\alpha$ has support on charges $q$ such that 
$q\times q=0$, and the last equation in \eqref{BPSCons} implies that a Fourier coefficient 
with $q\neq 0$ can only depend on the $E_{6(6)}/Sp(4)$ moduli through the invariant mass $|V(q)|$ (as was argued  on the basis of the Gelfand--Kirillov dimension)\footnote{While the three equations in \eqref{BPSCons} admit a solution of
the form $ F_\Gamma(R,t) = R^{10}F_\Gamma(t) $, this solution is not exponentially suppressed at large $R$ whereas Fourier coefficients are, and it does not extend to a solution of \eqref{E6}.}:
\be 
F_\Gamma(R,t) = F_{\Gamma,0}(R,\phi,y) + 
\sum_{q\in \IZ^{\scalebox{0.5}{$27$}}_* \atop q\times q=0}\, F_{\Gamma,q}(R,\phi,|V(q)|^2) e^{2\pi i \tr q \,  \alpha } \ .\label{Fansatz} 
\ee
For the constant term in this expansion,  \eqref{BPSCons} is automatically satisfied and \eqref{E6}Ê implies that $F_{\Gamma,0}(R,\phi,y) $ must be proportional to the minimal $E_6$ theta series $\mathcal{E}^{\scalebox{0.7}{$E_6$}}_{\scalebox{0.6}{$\frac32\Lambda_1$}}(y)$ or do not depend on $y$, in agreement with the result found in \eqref{d4r4largestrong}. 
We shall now prove that the Fourier coefficients $F_{\Gamma,q}$
with $q\neq 0$ necessarily vanish. 
For this purpose one computes, for $\zeta= |V(q)|^2$,
\bea \cD_{ij}{}^{kl} F(\zeta)  &=& \Scal{Ê2 V(q)_{ij} V(q)^{kl} -4 V(q)_{[i}{}^{[k}  V(q)_{j]}{}^{l]} - \frac{1}{6} \delta_{ij}^{kl} \zeta } F^\prime(\zeta) \nn \\
\cD_{ijpq} \cD^{klpq} F(\zeta)  &=& \frac{2}{3} \Scal{Ê V(q)_{ij} V(q)^{kl} + V(q)_{[i}{}^{[k}  V(q)_{j]}{}^{l]} + \frac{2}{9} \delta_{ij}^{kl} \zeta } \nn\\
&& \hspace{5mm}\times \Scal{Ê2 |V(q)|^2 F^{\prime\prime}Ê(\zeta) + 5  F^{\prime}Ê(\zeta) }
+ \frac{4}{3}  \delta_{ij}^{kl} \zeta\, F^{\prime}Ê(\zeta)  \ .  \label{DecE6D2}
\eea
The function \eqref{Fansatz} solves the two first equations of  \eqref{BPSCons} for a 1/2-BPS charge $q$, and the third gives
\be \Scal{ 4 \zeta\tfrac{\partial \, }{\partial \zeta} + 12 +  \partial_\phi}\scal{ R \tfrac{\partial \, }{\partial R}-10} F_{\Gamma,q}(R,\phi,\zeta) = 3 ( 2 \pi R e^{3\phi} n)^2    \tfrac{\partial \, }{\partial \zeta}F_{\Gamma,q}(R,\phi,\zeta) \ . \label{TensorEq1}  \ee
Equation \eqref{E6} decomposes into three equations associated to the three distinct tensor structures in \eqref{DecE6D2}
\bea \Scal{4 \zeta \tfrac{\partial \, }{\partial \zeta}+16 +\partial_\phi} \scal{  R \tfrac{\partial \, }{\partial R}-10}  \frac{\partial  F_{\Gamma,q}(R,\phi,\zeta)  }{\partial \zeta}&=& 3 ( 2 \pi R e^{3\phi} n)^2 \frac{\partial  F_{\Gamma,q}(R,\phi,\zeta)  }{\partial \zeta}  \nn \\
 \Scal{\scal{ -2 \zeta \tfrac{\partial \, }{\partial \zeta}+1 +\partial_\phi}\tfrac{\partial \, }{\partial \zeta}+ 3 (2\pi)^2 e^{-4\phi}} \scal{  R \tfrac{\partial \, }{\partial R}-10} F_{\Gamma,q}(R,\phi,\zeta) \hspace{-1mm} &=&\hspace{-1mm}   6 ( 2 \pi R e^{3\phi} n)^2   \frac{\partial  F_{\Gamma,q}(R,\phi,\zeta)  }{\partial \zeta} \nn \\\label{TensorEq2} 
\eea
 and 
 \begin{multline}
  \Scal{ 10 \zeta^2 \tfrac{\partial^2 \, }{\partial \zeta^2} + 67 \zeta  \tfrac{\partial \, }{\partial \zeta} - \zeta \tfrac{\partial \, }{\partial \zeta} \partial_\phi - 24 \partial_\phi - 2 \partial_\phi^{\; 2}+ 18 \pi^2 e^{-4\phi} \zeta }  \scal{  R \tfrac{\partial \, }{\partial R}-10} F_{\Gamma,q}(R,\phi,\zeta) \\ = -3 ( 2 \pi R e^{3\phi} n)^2    \scal{  3 R \tfrac{\partial \, }{\partial R}+\zeta \tfrac{\partial \, }{\partial \zeta}  - 18 + \partial_\phi} F_{\Gamma,q}(R,\phi,\zeta) \ . \label{TensorEq3}
  \end{multline} 
  Note that the first equation in \eqref{TensorEq2} is the derivative of \eqref{TensorEq1}, but the three remaining equations are independent. Moreover the Laplace equation \eqref{180Laplace}  decomposes as 
 \begin{multline}\Scal{ \tfrac{1}{4} R^2 \tfrac{\partial^2 \, }{\partial R^2} -  \tfrac{57}{4} R \tfrac{\partial \, }{\partial R} + \tfrac{1}{12} \partial_\phi^{\; 2}  + \tfrac92  \partial_\phi + \tfrac83 \zeta^2 \tfrac{\partial^2 \, }{\partial \zeta^2} + \tfrac{56}{3}\zeta \tfrac{\partial \, }{\partial \zeta}  }  F_{\Gamma,q}(R,\phi,\zeta)  \\
 = \Scal{(2\pi e^{-2\phi})^2 \zeta + ( 2\pi R e^{3\phi} n e^{3\phi})^2- 180 }  F_{\Gamma,q}(R,\phi,\zeta) \ .   \end{multline} 
As a Fourier coefficient,   $F_{\Gamma,q}(R,\phi,\zeta)$ must behave as $e^{-2\pi R e^{3\phi} n }$ as $R$ goes to infinity. Solving \eqref{TensorEq1} and \eqref{TensorEq2} perturbatively in the $\frac{1}{R}$ expansion one obtains the continuous family of solutions 
 \be  
 F_{\Gamma,q}(R,\phi,\zeta) = \frac{e^{-9\phi}}{\zeta^{\frac{3}{4}}} K_{\frac{3}{2}}(2\pi e^{-2\phi} \sqrt{\zeta}) 
 \int \de s\, \mu(s) \, R^{10+s} \,  e^{-3s\phi} \, K_s(2\pi R e^{3\phi} n) \ . 
 \ee
 However, there are no choice of $\mu(s)$ which would satisfy  either \eqref{TensorEq3} or the Laplace equation. One arrives at the same conclusion assuming that $F_{\Gamma,q}(R,\phi,\zeta)$ behaves as a power of $R$ as $R\to 0$ and solving  \eqref{TensorEq1} and \eqref{TensorEq2} perturbatively in the $R$ expansion. Therefore we conclude that although this type of Fourier coefficient is naively allowed on the basis of the Gelfand--Kirillov dimension, it is in fact ruled out by the tensorial differential equation associated to the next-to-minimal representation of $E_8$.

\section{Invariant lattices for exceptional Chevalley groups\label{sec_e8lat}}Ê

In this section we construct a $G(\IZ)$-invariant lattice $M_{\mathfrak{g}}$ in the Lie algebra of any 
split real simple group $G(\IR)$ obtained from a Jordan algebra of degree 3 via  Freudenthal's
triple system construction. Such a  construction exists for all split exceptional groups 
and for the split orthogonal groups of rank higher than 4 and $SO(4,3)$ (see {\it e.g.} table 1 of \cite{krutelevich2007jordan}). For definiteness, we consider the case $G=E_8$, which 
is based on the  Jordan algebra of 3 by 3 Hermitian matrices 
over the integral split octonions, but the cases $E_7$ and $E_6$ can be obtained by replacing
the split octonions by split quaternions and split complex numbers, respectively. In the following we
use the alternative notation $M_{\Lambda_8}^{E_8} \equiv M_{\mathfrak{e}_8}$,  $M_{\Lambda_1}^{E_7} \equiv M_{\mathfrak{e}_7}$,  $M_{\Lambda_2}^{E_6} \equiv M_{\mathfrak{e}_6}$, where
in each case $\Lambda_i$ is the fundamental weight corresponding to the adjoint representation 
of $G$. We also recall the definition of other lattices which enter in the construction of $M_{\mathfrak{g}}$.

We start with the 27-dimensional lattice $M^{E_6}_{\Lambda_1}$ of integral elements of the 
27-dimensional exceptional Jordan algebra $\cJ$ of 3 by 3 Hermitian matrices over the integral split octonions. The algebra $\cJ$ has automorphism group $F_{4(4)}$. $M^{E_6}_{\Lambda_1}$ admits a 
integer valued cubic form $I_3(q)=\det\vq$ and a  symmetric cross product  $q\mapsto q\times q$ 
valued in its dual $M^{E_6}_{\Lambda_6}\equiv (M^{E_6}_{\Lambda_1})^*$, normalized such that 
$\tr [\vq \times \vq \, \vq] = 3 \det \vq$. The continuous group $E_{6(6)}$ acts on 
$M^{E_6}_{\Lambda_1}$ and $M^{E_6}_{\Lambda_6}$ in the representations ${\bf 27}$ and $\overline{ \bf 27}$
with highest weight $\Lambda_1$ and $\Lambda_6$, respectively. The Chevalley group $E_6(\IZ)$ is defined as the subgroup of $E_{6(6)}$ which preserves $M^{E_6}_{\Lambda_1}$, or equivalently 
$M^{E_6}_{\Lambda_6}$. 

Next, we consider  the 56-dimensional lattice $M^{E_7}_{\Lambda_7}=
\IZ\oplus (M^{E_6}_{\Lambda_1})^* \oplus M^{E_6}_{\Lambda_1} \oplus \IZ$,  equipped with the integer 
symplectic pairing
\be
\label{defsymp}
\langle \Gamma, \Gamma' \rangle = q_0 p'^0 + \tr \vq \vp' -  \tr \vq' \vp - q'_0 p^0 \in \IZ\ ,
\ee
where $\Gamma=(q_0,\vq,\vp,p^0)$, $\Gamma'=(q'_0,\vq',\vp',p'^0)$,
and the integer valued quartic invariant
\be
I_4(\Gamma)=- 4 q_0\det\vp +4 p^0 \det\vq +4 \tr[\vp\times\vp\,\vq\times\vq]
    -(p^0 q_0 + \tr \vp \vq)^2 \,.
\label{I4-Albert}
\ee
Using Freudenthal's construction  \cite{freudenthal1955beziehungen}, one obtains the action of $E_{7(7)}$ on $M^{E_7}_{\Lambda_7}\otimes \IR$ in the representation ${\bf 56}$ with highest weight $\Lambda_7$,
which preserves both the symplectic pairing \eqref{defsymp} and the quartic invariant $I_4(\Gamma)$.
The Chevalley group $E_7(\IZ)$ is the subgroup of $E_{7(7)}$ which preserves the
lattice  $M^{E_7}_{\Lambda_7}$ \cite{integral-freud}.
For $\Gamma\in M^{E_7}_{\Lambda_7}$, the polynomial $I_4(\Gamma)$ is either a multiple of $4$, in which case we will say that $\Gamma$ is even, or $I_4(\Gamma)=-1$ mod $4$ in which case we will say that $\Gamma$ is odd. 
The gradient
of the quartic invariant  $I_4(\Gamma)$ defines a trilinear map $I_4^\prime(\Gamma)$, 
\be 
4 I_4(\Gamma) = \langle \Gamma , I_4^\prime(\Gamma) \rangle\ ,
\ee
such that
\be I_4^\prime(\Gamma,\Gamma,\Gamma) = 6 I_4^\prime(\Gamma) \ , 
\qquad
I_4^\prime( \Gamma , \Gamma , I_4^\prime(\Gamma)) = -8 I_4(\Gamma) \Gamma \ . 
\ee
For $\Gamma$ even, $I_4^\prime(\Gamma)$ is a multiple of $8$, while it is a multiple of $4$ for  $\Gamma$ odd. 
 These identities imply a few others that will be used in the following : \cite{Bossard:2013oga}
 \bea
  I_4^\prime(I_4^\prime(\Gamma) ) &=& -  16 [I_4(\Gamma)]^{2}\, \Gamma \ , \quad  I_4^\prime( \Gamma ,  I_4^\prime(\Gamma) , I_4^\prime(\Gamma)) = 8 I_4(\Gamma) I_4^\prime(\Gamma) \  , \nn \\
 I_4^\prime(\Gamma,\Gamma,I_4^\prime(\Gamma,\Gamma,\Upsilon) ) &=& - 16 I_4(\Gamma) \Upsilon - 8  \langle \Gamma , \Upsilon\rangle\, I_4^\prime(\Gamma) + 8 \langle  I_4^\prime(\Gamma)  , \Upsilon\rangle\,  \Gamma \ . 
  \label{I4ident}
\eea
The cross product of two charges $\Gamma_1\times\Gamma_2=\Gamma_2\times\Gamma_1$ is valued in the $\mathfrak{e}_{7(7)}$ algebra and is determined by its action on a vector $\Gamma$, 
\be  2( \Gamma_1 \times \Gamma_2) \cdot \Gamma =\frac{1}{4} I_4^\prime(\Gamma_1,\Gamma_2,\Gamma) 
-\tfrac{1}{2}Ê \langle \Gamma_1 , \Gamma \rangle\, \Gamma_2  
-\tfrac{1}{2}Ê \langle \Gamma_2 , \Gamma \rangle  \, \Gamma_1 
\ee
It satisfies the algebra \cite{freudenthal1955beziehungen}
\be [Ê\Gamma_1 \times \Gamma_2 , \Gamma_3 \times \Gamma_4] = ( ( \Gamma_1\times \Gamma_2) \cdot \Gamma_3 ) \times \Gamma_4 + \Gamma_3 \times ( ( \Gamma_1 \times \Gamma_2) \cdot \Gamma_4) \ , \ee
and the following identities 
\be I_4^\prime(\Gamma)\times  I_4^\prime(\Gamma) = 4 I_4(\Gamma)\, \Gamma \times \Gamma \ , \qquad \Gamma \times I'_4(\Gamma) = 0 \ . 
\ee
The cross product $\Gamma\times \Gamma$ is an integer if $\Gamma$ is even, and a half integer 
if it is odd. 

Finally, with a view towards the 5-graded decomposition \eqref{E8E7}, 
we consider the 248-dimensional lattice 
\be
M_{\mathfrak{e}_{8}} =
\IZ\oplus M^{E_7}_{\Lambda_7} \oplus M_0^{E_7} \oplus  M^{E_7}_{\Lambda_7}  \oplus \IZ\ ,
\ee
where $M_0$ is the 134-dimensional lattice of elements $(Q\in \mathfrak{e}_{7(7)}, \ell\in\IZ/2)$
such that $Q+\ell $ preserves $M^{E_7}_{\Lambda_7}$. 
$M_0^{E_7}$ is a reducible  $E_7(\IZ)$-module that decomposes as $M_0 \cong (M^{E_7}_{\Lambda_1},\IZ)\oplus (M^{E_7}_{\Lambda_1}+h(\Lambda_7),\IZ+\tfrac{1}{2}) $
 where $M^{E_7}_{\Lambda_1}$ is the lattice of elements in $\mathfrak{e}_7$ that preserve $M^{E_7}_{\Lambda_7}$, which is $M_{\mathfrak{e}_{7}}$ defined 
 by applying the same construction to the Jordan algebra of $3\times 3$ Hermitian matrices over the integral split quaternions,\footnote{This construction gives 
 $M_{\mathfrak{e}_{7}} =
\IZ\oplus M^{D_6}_{\Lambda_6} \oplus M_0^{D_6} \oplus  M^{D_6}_{\Lambda_6}  \oplus \IZ $ with $M_0^{D_6}=(M^{D_6}_{\Lambda_2},\IZ)\oplus (M^{D_6}_{\Lambda_2}+h(\Lambda_6),\IZ+\tfrac{1}{2})$, and $M^{D_6}_{\Lambda_2}$ is the lattice in $\mathfrak{so(6,6)}$ which preserves 
the three lattices  $ M^{D_6}_{\Lambda_6} $, $ M^{D_6}_{\Lambda_1}  $, $ M^{D_6}_{\Lambda_5}  $. Using the decomposition in $SL(6,\IZ)$ modules, 
$M^{D_6}_{\Lambda_6} \cong \IZ^\ord{-3} \oplus (\IZ^{15})^\ord{-1} \oplus   (\IZ^{15})^\ord{1}\oplus\IZ^\ord{3}$, $M^{D_6}_{\Lambda_1} \cong  (\IZ^{6})^\ord{-1} \oplus   (\IZ^{6})^\ord{1}$ and $M^{D_6}_{\Lambda_5} \cong (\IZ^6)^\ord{-2} \oplus (\IZ^{20})^\ord{0} \oplus   (\IZ^{6})^\ord{2}$, it follows that the elements of $M_{\mathfrak{e}_{7}}$ are precisely the ones that preserve $M^{E_7}_{\Lambda_7} = M^{D_6}_{\Lambda_1} \oplus M^{D_6}_{\Lambda_5} \oplus M^{D_6}_{\Lambda_1}$  since $h(\Lambda_6)\pm\frac12$ preserves $M^{D_6}_{\Lambda_1}$ and $h(\Lambda_6)$ preserves $M^{D_6}_{\Lambda_5}$.} and $h(\Lambda_7)$ is the element of  a chosen Cartan subalgebra of $\mathfrak{e}_{7(7)}$ associated to the weight $\Lambda_7$.\footnote{One could just as well
 choose the  Cartan generators $h(\Lambda_2),h(\Lambda_5)$, or any  element of the form $\Gamma\times \Gamma$ for any odd vector $\Gamma$, since all these elements are equivalent
 modulo  $M^{E_7}_{\Lambda_1}$.}  

We denote  elements of $M^{E_8}_{\Lambda_8}\equiv M_{\mathfrak{e}_{8}}$ 
 by $\mathcal{Q} = ( n,\Upsilon,Q+\ell,\Gamma,m)$,
where $n,m$ are integers, $\Upsilon,\, \Gamma$ elements of $\IZ^{56}$ and $(Q,\ell)\in M_0$.
An element $a\oplus b$ in the  nilpotent subalgebra $ {\bf 56}^{\ord{1}} \oplus {\bf 1}^{\ord{2}}$, defines the action of the `spectral flow' unipotent subgroup  on $\cQ$,
\bea 
m&\rightarrow& m + \langle a,\Gamma+b \Upsilon \rangle + 2 b \ell + b^2 n +\tfrac{1}{2}Ê \langle a , Q\cdot a \rangle +\tfrac{1}{4}  \langle \Upsilon , I_4^\prime(a) \rangle +\tfrac{1}{4}  n I_4(a) \ , \nn\\
\Gamma&\rightarrow&  \Gamma + Q\cdot a + \ell a + \tfrac{1}{8} I_4^\prime(a,a,\Upsilon) +\tfrac{1}{2}Êa  \langle a , \Upsilon \rangle  +\tfrac{1}{4} n  I_4^\prime(a) + b ( \Upsilon + a n )  \ , \nn\\
Q&\rightarrow&Q +  2 a \times \Upsilon +a \times a \, n \ , \qquad \ell \rightarrow \ell + \tfrac{1}{2}Ê\langle a , \Upsilon \rangle + b n   \ , \nn\\
\Upsilon&\rightarrow& \Upsilon +a n  \ , \nn\\
n&\rightarrow & n \  ,\label{Taction}
\eea
while an element $\bar a \oplus \bar b$ in the opposite nilpotent subalgebra ${\bf 1}^{\ord{-2}} \oplus {\bf 56}^{\ord{-1}}$ acts via
\bea m&\rightarrow& m \ , \nn\\
\Gamma&\rightarrow& \Gamma - \bar a m \ , \nn\\
Q&\rightarrow&Q - 2 \bar a \times \Gamma +Ê\bar a \times \bar a m \ , \qquad \ell \rightarrow \ell +\tfrac{1}{2}Ê \langle \bar a , \Gamma \rangle - \bar b m   \ , \nn\\
\Upsilon&\rightarrow& \Upsilon   + Q\cdot \bar a - \ell \bar a - \tfrac{1}{8} I_4^\prime(\bar a,\bar a,\Gamma) +\tfrac{1}{2}Ê \langle \Gamma , \bar a\rangle \bar a +\tfrac{1}{4}   m I_4^\prime(\bar a)  - \bar b ( \Gamma - \bar a m )  \ , \nn\\
n&\rightarrow & n +  \langle \bar a,\Upsilon-\bar b \Gamma \rangle - 2\bar  b \ell + {\bar b}^2 m +\tfrac{1}{2}Ê \langle \bar a , Q\cdot \bar a \rangle - \tfrac{1}{4} \langle \Gamma , I_4^\prime(\bar a) \rangle + \tfrac{1}{4}  m I_4(\bar a)\  . \label{CTaction}
\eea
The Chevalley group $E_8(\IZ)$ is generated by the Chevalley group $E_7(\IZ)$ and the unipotent generator defined above with $b\in \IZ$ and $a\in \IZ^{56}$ such that $a\times a=0$. More generally all the elements \eqref{Taction} such that $a$ is even and $b\in \IZ$, or such that $a$ is odd and $b\in \IZ+\tfrac{1}{2}$ are in $E_8(\IZ)$, and similarly for \eqref{CTaction}.


\providecommand{\href}[2]{#2}\begingroup\raggedright\endgroup

\end{document}